\documentclass[pre,longbibliography,superscriptaddress]{revtex4-1}

\usepackage{graphics}
\usepackage{graphicx}
\usepackage{amsmath}
\usepackage{amssymb,color}

\newcommand{\textin}[1]{\mbox{\scriptsize{#1}}}

\definecolor{grisclair}{rgb}{0.6,0.6,0.6}

\newcommand{\blue}[1]{\textcolor{blue}{#1}}

\newcommand{\beq}{\begin{equation}}
\newcommand{\ee}{\end{equation}}

\begin{document}

\title{\blue{Breakup of an electrified viscoelastic liquid bridge}}
\author{Manuel Rubio}
\address{Departamento de Ingenier\'{\i}a Mec\'anica, Energ\'etica y de los Materiales and\\
Instituto de Computaci\'on Cient\'{\i}fica Avanzada (ICCAEx),\\
Universidad de Extremadura, Avda.\ de Elvas s/n, E-06071 Badajoz, Spain}
\author{Emilio J. Vega}
\address{Departmento de Ingenier\'{\i}a Mec\'anica, Energ\'etica y de los Materiales and\\
Instituto de Computaci\'on Cient\'{\i}fica Avanzada (ICCAEx),\\
Universidad de Extremadura, Avda.\ de Elvas s/n, E-06071 Badajoz, Spain}
\author{Miguel A. Herrada}
\address{Departamento de Mec\'anica de Fluidos e Ingenier\'{\i}a Aeroespacial,\\
Universidad de Sevilla, E-41092 Sevilla, Spain}
\author{Jos\'e M. Montanero}
\address{Departmento de Ingenier\'{\i}a Mec\'anica, Energ\'etica y de los Materiales and\\
Instituto de Computaci\'on Cient\'{\i}fica Avanzada (ICCAEx),\\
Universidad de Extremadura, Avda.\ de Elvas s/n, E-06071 Badajoz, Spain}
\author{Francisco J. Galindo-Rosales}
\address{CEFT, Departamento de Engenharia Qu\'{\i}mica,\\
Faculdade de Engenharia da Universidade do Porto,\\
Rua Dr.\ Roberto Frias, 4200-465 Porto, Portugal}

\begin{abstract}
We study both numerically and experimentally the breakup of a viscoelastic liquid bridge formed between two parallel electrodes. The polymer solutions and applied voltages are those commonly used in electrospinning and near-field electrospinning. We solve the leaky-dielectric FENE-P model to describe the dynamical response of the liquid bridge under isothermal conditions. The results show that the surface charge screens the inner electric field perpendicular to the free surface over the entire dynamical process. The liquid bridge deformation produces a normal electric field on the outer side of the free surface that is commensurate with the axial one. The surface conduction does not significantly affect the current intensity in the time interval analyzed in the experiments. The force due to the shear electric stress becomes comparable to both the viscoelastic and surface tension forces in the last stage of the filament. However, it does not alter the elasto-capillary balance in the filament. As a consequence, the extensional relaxation times measured from the filament exponential thinning approximately coincides with the stress relaxation time prescribed in the FENE-P model. The above results allow us to interpret correctly the experiments. In the experiments, we measure the filament electrical conductivity and extensional relaxation time for polyethylene oxide (PEO) dissolved in deionized water and in a mixture of water and glycerine. We compare the filament electrical conductivity with the value measured in hydrostatic conditions for the same estimated temperature. Good agreement was found for PEO dissolved in water+glycerine, which indicates that the change of the filament microscopic structure due to the presence of stretched polymeric chains does not significantly alter the ion mobility in the stretching direction. Significant deviations are found for PEO dissolved in deionized water. These deviations may be attributed to the heat transferred to the ambient, which is neglected in the calculation of the filament temperature. We measure the extensional relaxation time from the images acquired during the filament thinning. The relaxation times obtained in the first stage of the exponential thinning hardly depend on the applied voltage. Little but measurable influence of the applied voltage is found in the last phase of the filament thinning.
\end{abstract}
\draft
%\pacs{47.55.Dz, 47.20.Cq, 47.15Hg,47.55Bx}
\date{\today}
\maketitle

\section{Introduction}
\label{sec1}

% Electric field and viscoelasticity
Electric fields and viscoelasticity play a fundamental role in a multitude of natural processes and technological applications. In many situations, these two factors come into play simultaneously to determine the evolution of the system. A very important example of this is electrospinning \citep{F02,CJ06}, in which a viscoelastic microjet is ejected from a thin feeding capillary by the action of an externally applied electric field. The subsequent solidification of the jet results in the extrusion of fibers with diameters down to the nanometer scale. In traditional electrospinning, fibers are produced chaotically. This feature limits the application of electrospinning in devices that demand arranged or patterned micro/nanoscale fibrous structures. Ultrafine fibers with unique physical and chemical properties can be deposited with near-field electrospinning in a direct, continuous, and controllable manner \citep{SCLL06}. These fibers can be used in electronic components, flexible sensors, energy harvesting, and tissue engineering, among other applications \citep{HZYYYNL17}. Sophisticated versions of this technique are continuously emerging \citep{LRC20}. Despite the importance of technologies such as electrospinning \citep{RYFK00,YKR01,RY08,YPR14} and near-field electrospinning \citep{SCLL06}, the interaction between the applied electric field and liquid viscoelasticity is not well understood.

% The leaky-dielectric approximation
When a viscoelastic thread is stretched by hydrodynamic \citep{ZF10b,PVCM17,HKHSTF18,PORRVM19,VKVKKMT19} or electrical forces \citep{F02,CJ06,RYFK00,YKR01,RY08,YPR14}, the polymers dissolved in the liquid uncoil. If the flow strain rate is sufficiently large in terms of the inverse of the polymer longest relaxation time, then the coiling-to-stretching transition takes place, and the extensional viscosity grows in time until polymers are fully stretched \citep{EH97,EV08}. In this situation, the extensional viscosity becomes one of the dominant factors of the liquid flow. On the other hand, interfaces between immiscible phases form barriers that prevent the continuous diffusion of free ions under applied electric fields, which causes the accumulation of free charge onto those interfaces. If the electric relaxation time is much smaller than any characteristic time of the system, then free charge in the bulk is assumed to be zero, which constitutes the essential approximation in the so-called leaky-dielectric model \citep{T66,MT69,S97,GLHRM18}. The existence of charge on the interface gives rise to tangential Maxwell stresses. These stresses also appear at the interface owing to the difference between electrical permittivities of the adjacent media.

% The bulk electrical conductivity
One of the key elements in the description of phenomena like electrospinning is the modeling of the diffusion of free ions across the liquid bulk. It is typically assumed that ions distribute uniformly for distances from the interface much larger than the thickness of the the diffuse Debye layer \citep{T66,MT69,S97,GLHRM18}. In most applications, this thickness is several orders of magnitude smaller than the bulk size, which invites one to regard the bulk as an Ohmic medium of constant scalar conductivity. However, one may wonder whether the presence of macromolecules significantly stretched may render electric conduction anisotropic in the bulk \citep{BHGM19}. Specifically, the conductivity in some direction may be significantly different from that measured under hydrostatic conditions because the polymer stretching may facilitate/hinder the diffusion of ions along that direction. To the best of our knowledge, the influence of the stretching polymers on the ohmic conduction across the polymer solution has not been considered yet. Previous studies have derived scaling laws for the total electric current transported by the jet in electrospinning \citep{BSBMR10}, establishing the differences with respect to those for electrospray, and in near-field electrospinning \citep{CTWFL10}.

% Surface conduction
The charge flux into the interface can be balanced by, among other mechanisms, the lateral conduction across the Debye layer formed next to that interface. On a macroscopic scale, the net free charge accumulated in the Debye layer is quantified by the surface charge concentration, and the lateral conduction can be represented with a surface conductivity. This conductivity is expected to differ significantly from that in the bulk due to the electrical structure of the diffusive Debye layer, which means that charge transport near the interface becomes anisotropic. There is no consensus about the dependency of the surface conductivity upon parameters such as the surface charge concentration. While some authors consider a constant surface conductivity \citep{BS02,BT11}, others assume that the ion mobility in the Debye layer equals that in the bulk \citep{GRGL20}, which results into a surface conductivity proportional to the surface charge density. For fluid configurations characterized by large surface-to-volume ratios, such as electrospinning, surface conduction may in principle become comparable to or even dominate over Ohmic conduction in the bulk.

% The extensional relaxation time
Weakly viscoelastic polymer solutions with quasi-monodisperse molecular weight distributions are commonly used in applications such as electrospinning. The elastic properties of these solutions can be approximately quantified by a single characteristic relaxation time as long as the relaxation time of the entire chain is much smaller than that of the subchains \citep{CPKOMSVM06}. The experimental determination of this relaxation time is generally challenging both in shear and extensional flows \citep{DHS17}. Conventional rotational rheometers cannot provide reliable results when dealing with viscoelastic liquids with low elasticity and low viscosity mainly due to the onset of inertial instabilities \citep{EJC15}. The development of extensional flow-based rheometry has experienced a certain delay with respect to that of its shear counterpart \citep{D17}. However, extensional rheometry has demonstrated to be a more successful approach for determining the relaxation time of dilute polymer solutions both at macro- \citep{VMC12, KSHKMCTM15, DZJS15, SVSMA17} and micro-scales \citep{OC06,CGPAO11, ZSWPAL14, GOA14,HOAM12, H16}. Filament thinning rheometers such as CaBER \citep{MT00} and FiSER \citep{AM01} have been considered as accurate devices for the characterization of viscoelastic fluids over the last twenty years, being able to produce a quasi-ideal uniaxial extensional deformation \citep{A01}.

% Rheometry + electric fields
The above comments refer to the determination of the relaxation time without the application of an external electric field. Rotational rheometers can be equipped with an electrorheological cell, which allows for the application of an external electric field perpendicular to the flow direction. Typically, the voltage difference between two solid surfaces containing the sample is limited to current intensities of the order of tens of $\mu$A. For that reason, the electrorheological properties of leaky-dielectric viscoelastic fluids under simple shear flow have not been reported in the literature yet. Neither CaBER \citep{MT00} nor FiSER \citep{AM01} extensional rheometers are currently commercialized with an electrorheological cell to apply an electric field while the fluid sample undergoes the rheological characterization. Very recently \citep{SNG20, GSG19}, Sadek et al. have designed, prototyped and validated a new electrorheological cell for the CaBER device, which allows imposing an electric field aligned with the direction of the flow. This configuration is very convenient for the electrorheological characterization of viscoelastic fluids because the electric and velocity fields are parallel to each other, as occurs in electrospinning. However, and as mentioned above, the current intensity in the high voltage power supply is typically limited to tens of $\mu$A, which has made impossible the study of the effects of electric fields commonly applied in electrospinning.

% Theoretical analysis
Accurate codes for studying viscoelastic flows have been developed over the last three decades (see, e.g., \citep{BABY86,PS02}). Simulations have been conducted to examine capillary-driven viscoelastic flows in liquid bridges (see, e.g., \citep{BBP08,BAHPMB10,LRP13,EHS20}). Powerfull numerical methods have also been designed to study electrohydrodynamic phenomena in  leaky-dielectric liquids with free surfaces \citep{CJHB08}. Electrohydrodynamic phenomena in leaky-dielectric liquid bridges have been analyzed numerically in some occasions too \citep{PET01,BS02,MP20}. However, the combined action of viscoelasticity and electric fields on the capillary breakup of a liquid bridge has not as yet been considered. In this paper, we will numerically study the capillary instability arising in a liquid bridge of a viscoelastic liquid subject to an axial electric field by solving the leaky-dielectric \citep{S97,GLHRM18} FENE-P \citep{BAH87,EH97,EV08} model. This analysis not only constitutes a novel contribution but also allows us to gain insight into the interplay between inertia, surface tension, viscoelasticity, and Maxwell stresses in the experiments conducted in this work. The numerical study will be conducted under isothermal conditions, which will help us to study the influence of the Maxwell stresses alone, excluding the thermal effects associated with the Joule heating inherent to the problem. This will enable us to interpret correctly our experimental results.

% Experimental analysis
In this paper, we will analyze experimentally the breakup of an axisymmetric liquid bridge formed between two disks kept at different voltages. The upper disk will be moved at a small speed away from the lower one until the maximum liquid bridge length is reached \citep{CC10}. The capillary instability triggered in that limit gives rise to the formation of a quasi-cylindrical liquid filament with a diameter much smaller than that of the supporting disks. The synchronized measurement of the current intensity across that filament and its diameter will allow us to describe the phenomenon. We will consider both polymer solutions and voltages typically used in electrospinning and near-field electrospinning.

\section{Mathematical model}
\label{sec2}

\subsection{The leaky-dielectric FENE-P model}

\begin{figure}
\begin{center}
\includegraphics[width=0.4\linewidth]{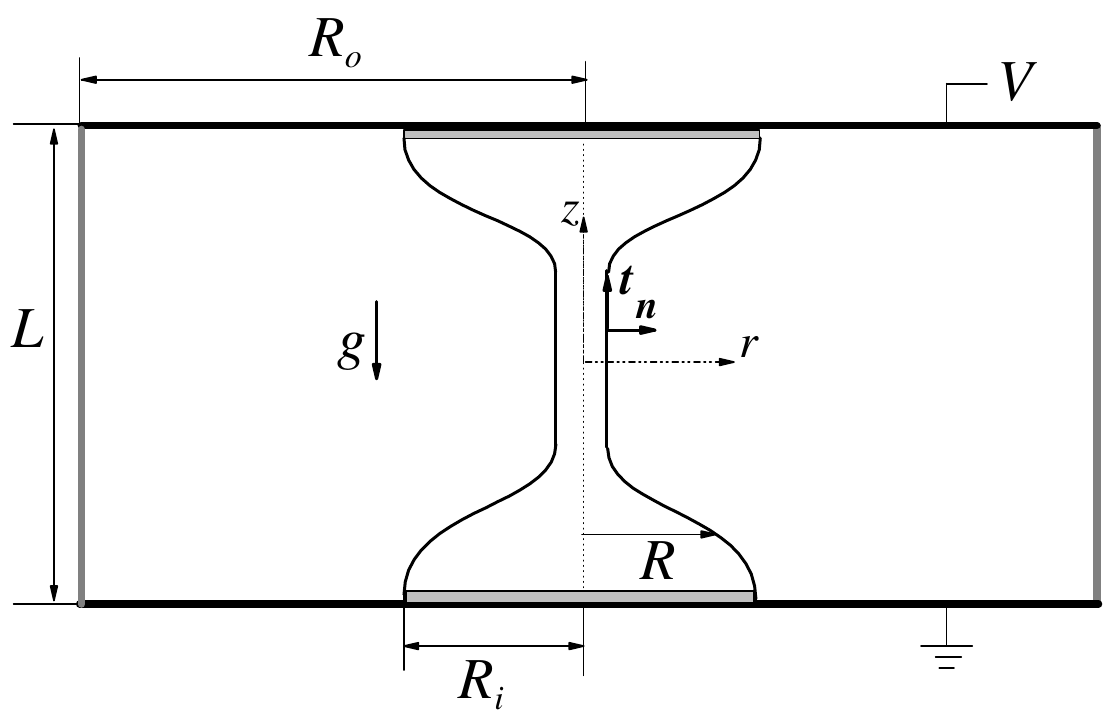}
\end{center}
\caption{Sketch of the problem's formulation}
\label{sketch2}
\end{figure}

% Liquid bridge
In the theoretical study, we consider a liquid bridge of constant volume ${\cal V}$ and density $\rho$. The liquid bridge is surrounded by a dielectric fluid of negligible density and viscosity, and is subjected to the action of the gravity $g$ in the axial direction (Fig.\ \ref{sketch2}). The liquid bridge is held by the constant surface tension $\gamma$ between two horizontal electrodes of radius $R_o$ and separated by a distance $L$. A constant voltage drop $V_0$ is applied between the two electrodes. The triple contact lines are pinned to the solid surfaces at a distance $R_i$ ($R_i\ll R_o$) from the liquid bridge axis. In addition, $K$ is the liquid bridge electrical conductivity, and $\varepsilon_0$ and $\varepsilon=\beta \varepsilon_0$ are the outer medium and liquid bridge electrical permittivities, respectively. 

% Surface conductivity
Free net charge is accumulated within the Debye layer formed on the inner side of the liquid-fluid interface. We assume that conduction across that layer is essentially caused by the excess of charge. For water droplets, the excess of charge at the interface is essentially due to either H$^+$ or OH$^-$ (adsorbed impurities from the atmosphere that are ionized at the free surface are assumed to play a secondary role because their mobilities are much smaller than those of H$^+$ and OH$^-$). The electric drift $v_e$ of these ions can be calculated as $v_e=\lambda\, E_t$, where $\lambda$ is the charge mobility multiplied by the valence, and $E_t$ is the electric field tangential to the interface \citep{GLHRM18,GRGL20}. We assume that the charge mobility in the Debye layer equals the mean value of that of H$^+$ and OH$^-$ in the bulk.

% Dimensionless quantities
In this section, all the quantities are made dimensionless with the triple contact line radius $R_i$, the liquid density $\rho$, the surface tension $\gamma$, and the voltage drop $V_0$. This choice yields the characteristic time, velocity, pressure and electric field scales $t_c = (\rho R_i^3/\gamma)^{1/2}$, $v_c =R_i/t_c$, $p_c=\gamma/R_i$ and $E_c=V_0/R_i$, respectively. For the sake of simplicity, in the rest of this section the symbols represent dimensionless quantities, while they stand for the dimensional counterparts in the rest of the paper.

% Navier-Stokes equations
The velocity ${\bf v}(r,z,t)=u(r,z,t){\bf e_r}+w(r,z,t){\bf e_z}$ and modified pressure (the hydrostatic pressure plus gravitational potential per unit volume) $p(r,z,t)$ fields are calculated from the continuity and momentum equations
\begin{equation}
\label{e1}
{\boldsymbol \nabla}\cdot {\bf v}=0, 
\end{equation}
\begin{equation}
\label{m}
\frac{\partial {\bf v}}{\partial t}+{\bf v}\cdot {\boldsymbol \nabla}{\bf v}=-{\boldsymbol \nabla}p+{\boldsymbol \nabla}\cdot {\bf T}.
\end{equation}
The extra stress tensor ${\bf T}$ in the FENE-P model \citep{BAH87,EH97,EV08} can be seen as the sum of the solvent contribution ${\bf T}_s=\text{Oh}_s [{\boldsymbol \nabla}{\bf v}+({\boldsymbol \nabla}{\bf v})^T]$ and that due to the presence of polymers 
\begin{equation}
\label{h10}
{\bf T}_p=\frac{\text{Oh}_0}{\lambda_s} \left(1-\frac{\eta^{(s)}}{\eta_0}\right) {\cal F}(I_1) ({\bf A}-{\bf I}), \quad {\cal F}(I_1)=\frac{L^2-3}{L^2-I_1}, \quad I_1=\text{tr}({\bf A})\; .
\end{equation}
This stress relationship is complemented by the nonlinear relaxation law
\begin{equation}
{\stackrel{\triangledown}{\mathbf A}}=-\frac{1}{\lambda_s}({\cal F}(I_1) {\bf A}-{\bf I}).
\end{equation}
In the above equations, Oh$_s=\eta^{(s)}(\rho R_i \gamma)^{-1/2}$ and Oh$_0=\eta_0(\rho R_i \gamma)^{-1/2}$ are the Ohnesorge numbers defined in terms of the solvent viscosity  $\eta^{(s)}$ and solution zero-shear viscosity $\eta_0$, respectively, $\lambda_s$ is the dimensionless stress relaxation time (the Deborah number), $L^2$ is the finite extensibility parameter, and ${\stackrel{\triangledown}{\mathbf A}}$ is the upper-convected time derivative of the conformation tensor $\mathbf{A}$.

In the leaky-dielectric model, the bulk net free charge is assumed to be negligible, and, therefore, the electric potentials $\phi^i$ and $\phi^o$ in the inner and outer domains obey the Laplace equation 
\begin{equation}
{\boldsymbol \nabla}^2\phi^{i,o}=0\label{el1}.
\end{equation}
The inner and outer electric fields ${\bf E}^{i,o}= E^{i,o}_r {\bf e_r}+ E^{i,o}_z {\bf e_z}$ are calculated as ${\bf E}^{i,o}={\boldsymbol \nabla} \phi ^{i,o}$.

% Interface boundary conditions
The free surface location is defined by the equation $r=R(z,t)$. The boundary conditions at that surface are:
\begin{equation}
\frac{\partial R}{\partial t}+R_z w-u=0,\label{int1}
\end{equation}
\begin{equation}
-p+Bz-\frac{RR_{zz}-1-R_z^{2}}{R(1+R_z^{2})^{3/2}}+{\bf n}\cdot {\bf T}\cdot {\bf n}=\frac{\chi}{2}\left[(E_n^o)^2-\beta (E_n^i)^2\right]+\chi\frac{\beta-1}{2}(E_t)^2,\label{int3}
\end{equation}
\begin{equation}
{\bf t}\cdot {\bf T}\cdot {\bf n}=\sigma E_t, \label{int2}
\end{equation}
where $R_z\equiv dR/dz$ and $R_{zz}=d^2R/dz^2$, $B=\rho g R_i^2/\gamma$ is the gravitational Bond number, ${\bf n}$ is the unit outward normal vector, $\chi=\varepsilon_o V_0^2/(R_i\gamma)$ is the electric Bond number, ${\bf t}$ is the unit vector tangential to the free surface meridians, and $\sigma$ is the surface charge density. Equation (\ref{int1}) is the kinematic compatibility condition, while Eqs.\ (\ref{int3}) and (\ref{int2}) express the balance of normal and tangential stresses on the two sides of the free surface, respectively. The right-hand sides of these equations are the Maxwell stresses resulting from both the accumulation of free electric charges at the interface and the jump of permittivity across that surface. The pressure in the outer medium has been set to zero. 

The electric field at the free surface and the surface charge density are calculated as
\begin{equation}
E_n^i=\frac{-R_z E^i_{z}+ E^i_r}{\sqrt{1+R_z^2}}, \quad E_n^o=\frac{-R_zE^o_z+ E^o_r}{\sqrt{1+R_z^2}} \label{int4},
\end{equation}
\begin{equation}
E_t= \frac{R_z E^o_r+E^o_z}{\sqrt{1+R_z^2}}=\frac{R_zE^i_r+E^i_z}{\sqrt{1+R_z^2}}, \label{nose}
\end{equation}
\begin{equation}
\sigma=\chi(E_n^o-\beta E_n^i),\label{int6}
\end{equation}
It must be noted that the continuity of the electric potential across the free surface, $\phi^i=\phi^o$, has been considered in Eq.\ (\ref{nose}).

The free surface equations are completed by imposing the surface charge conservation at $r=R(z,t)$,
\begin{equation}
\frac{\partial \sigma}{\partial t}+\sigma v_n({\boldsymbol \nabla}\cdot {\bf n})+\boldsymbol{\nabla_s}\cdot (\sigma{\bf v_s})+\lambda \boldsymbol{\nabla_s}\cdot (\sigma E_t {\bf t})=\chi \alpha E_n^i,
\label{int7}
\end{equation}
where $\boldsymbol{\nabla_s}$ is the tangential intrinsic gradient along the free surface, ${\bf v_s}=v_t {\bf t}$ is the projection of the velocity of a free surface element onto the free surface, and $\alpha=K\left[\rho R_i^3/(\gamma \varepsilon_o^2)\right]^{1/2}$ is the dimensionless electrical conductivity. The diffusion term has been neglected because it is usually much smaller than the dominant terms \citep{GLHRM18}.

% Other boundary conditions
The anchorage condition $R=1$ is set at $z=\pm \Lambda$, where $\Lambda=L/(2R_i)$ is the slenderness. The nonslip boundary condition is imposed at the solid surfaces in contact with the liquid. The nondimensional volume $\hat{{\cal V}}={\cal V}/(\pi R_0^2L)$ of the initial configuration is prescribed (and conserved), namely,
\begin{equation}
\int_{-\Lambda}^{\Lambda} R^2\ dz=2\Lambda \hat{{\cal V}}.\label{volume}
\end{equation}
The surface charge conservation equation (\ref{int7}) is integrated by assuming zero surface charge flux at the triple contact lines, and the regularity conditions $E^i_ r = u=w_r =0$ are prescribed on the symmetry axis. We fix the electric potential $\phi^{i,o}=0$ and $\phi_0$ at the lower and upper electrodes, respectively. The linear relationship $\phi^{o}=\phi_0 (z+\Lambda)/(2\Lambda)$ is set at the cylindrical lateral surface $r=R_o$.

% The initial condition
We start the simulation from a non-electrified liquid bridge at equilibrium with a slenderness just below the critical one. At the initial instant, we trigger the breakup process by applying a very small gravitational force (i.e., by slightly changing the Bond number value). As also done in the experiments, the voltage drop is applied at some instant before the elasto-capillary regime. Then, we simulate the liquid bridge breakup under the action of the electric field. 

% The current
The simulation allows one to calculate the total electric current $I$ as the sum of the contributions due to the bulk conduction $I_b$, surface convection $I_s^{(\textin{cv})}$, and surface conduction $I_{s}^{(\textin{cd})}$. These contributions can be calculated at any axial position $z$ along the liquid bridge as
\begin{equation}
\label{current}
I_b(z)=2 \pi \alpha \chi \int_0^{R(z)}  E^i_{z} (r,z)rd r,\quad I_s^{(\textin{cv})}(z)=2 \pi R(z) \sigma(z) v_{t}(z), \quad  I_s^{(\textin{cd})}(z)=2 \pi R(z) \sigma(z) \lambda E_t(z).
\end{equation}

%The 1D model
The theoretical model described above can be greatly simplified when the fluid adopts a slender shape along the streamwise direction $z$. In this case, the balance of forces becomes \citep{E97,G99b,F02,F03,CJ06}
\begin{eqnarray}
\label{mom}
\frac{\partial w}{\partial t}+w\, (w)_z=-B+\underbrace{\frac{(T_e)_z}{\pi R^2}}_{\text{\normalsize TE}}-\underbrace{\left(\frac{1}{R}\right)_z}_{\text{\normalsize ST}}+
\underbrace{\frac{\chi}{2}\left[(E_n^o)^2-\beta(E_n^i)^2\right]_z}_{\text{\normalsize SC}}+
\underbrace{\chi\frac{\beta-1}{2}\left[(E_t)^2\right]_z}_{\text{\normalsize PO}}+
\underbrace{\frac{2\sigma E_t}{R}}_{\text{\normalsize SE}},\nonumber\\
\end{eqnarray}
where $T_e=\pi R^2 (T_{zz}-T_{rr})$ is the tensile force in the filament due to both the solvent viscosity and polymeric stress, and the symbols $()_z$ and $[]_z$ indicate the derivative with respect to the $z$ coordinate. All the variables in Eq.\ (\ref{mom}) are supposed to be functions of the time $t$ and axial coordinate $z$ exclusively. To reduce the numerical noise in $T_e$, they are evaluated at the liquid bridge axis $r=0$ \citep{EHS20}. The labels indicate the nature of the corresponding term: tensile (TE) force, surface tension (ST) force, and electric force due surface charge (SC), polarization (PO) and shear electric (SE) stress. 

In the drops delimiting the filament, $1/R$ and $T_e$ take values much smaller than those in the filament. Assuming that the final stage of the filament thinning is dominated by surface tension and polymeric stress, the spatial integration of (\ref{mom}) from the upper/lower drop yields the force balance
\begin{equation}
\label{intmom}
\pi R+T_e=\widehat{T}_e(t),
\end{equation}
where $\widehat{T}_e(t)$ is the tension in the filament.

\subsection{Numerical method}

The leaky-dielectric FENE-P model was solved with a variation of the method described by \citet{HM16a}. The spatial physical domains occupied by the liquid and the outer dielectric medium were mapped onto two rectangular domains by means of the coordinate transformation. Each variable and its spatial and temporal derivatives appearing in the transformed equations were written as a single symbolic vector. Then, we used a symbolic toolbox to calculate the analytical Jacobians of all the equations with respect to the symbolic vector. Using these analytical Jacobians, we generated functions which could be evaluated in the course of the iterations at each point of the discretized numerical domains. 

The transformed spatial domains were discretized using $n_\eta^{(i)}=11$ and $n_\eta^{(o)}=35$ Chebyshev spectral collocation points \citep{KMA89} in the transformed radial direction $\eta$ of the inner and outer domains, respectively, as well as $n_\xi=251$ equally spaced collocation points in the transformed axial direction $\xi$. We increased the number of points in the axial direction up to $n_\xi=1001$ at the beginning of the filament thinning, when the polymeric stress blows up. The axial direction was discretized using fourth-order finite differences. Second-order backward finite differences were used to discretize the time domain. The time step was $\Delta t=0.01$. The non-linear system of discretized equations was solved at each time step using the Newton method. The method is fully implicit. During the exponential thinning, the rate $R^{-1} dR/dt$ of the relative variation of the filament radius remains practically constant. In this sense, the time scale of the process becomes fixed. If $\Delta t$ is much smaller than that time scale (as occurs in our simulation), the simulation remains resolved as the filament thins. We verified that the results did not change when the parameters of the initial grid were replaced with $\{n_\eta^{(i)}=13$,$n_\eta^{(o)}=35$,$n_\xi=351\}$, and the time step was reduced to $\Delta t=0.005$.

\section{Experimental method}
\label{sec3}

\subsection{Experimental setup and procedure}

\begin{figure}
\begin{center}
\includegraphics[width=0.3\linewidth]{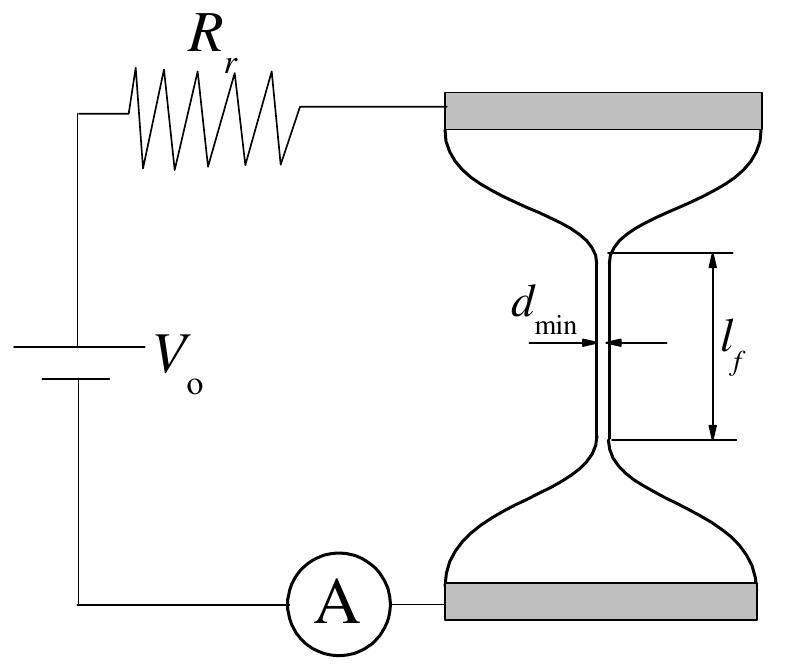}
\end{center}
\caption{Sketch of the experimental configuration}
\label{sketch}
\end{figure}

% Experimental setup
Figure \ref{sketch} shows a sketch of the experimental setup used in this work. A liquid bridge was held by the surface tension between two horizontal disks 2 mm in radius. The triple contact lines were pinned to the edges of those disks. The lower disk remained still, while the upper one was moved up at a constant speed using the capillary breakup extensional rheometer (HAAKE CaBER 1). A constant electric potential was applied during the last phase of the liquid bridge stretching with a DC high voltage power supply ({\sc LabSmith} HVS448). To limit the intensity current crossing the circuit and the voltage drop in the liquid bridge, a resistor was connected in series with the liquid bridge. The intensity current was measured with a picoammeter ({\sc Keithley} model 6485). Digital images of the liquid bridge were acquired at 1000-8000 frames per second (1028 x 1024 square pixels) depending on the experiment with a high-speed camera ({\sc Photron FASTCAM Mini UX100)}). The camera was equipped with a set of optical lenses (Optem Zoom 70 XL) with a variable magnification from 1$\times$ to 5:5$\times$. The liquid bridge was illuminated from the backside with white light provided by an optical fiber connected to a metal halide light source (LeicaEL6000). The optical fiber was connected to a 60mm Telecentric Backlight Illuminator (TECHSPEC), providing a truly collimated light and producing high contrast, silhouetted images. The power supply triggered the camera, which in turn triggered the picoammeter, so that the three devices were synchronized. 

% Experimental procedure
In the experiments, a liquid bridge of volume 25 mm$^3$ was formed between the supporting disks separated initially by a distance around 2 mm. The liquid bridge was stretched by moving the upper disk away from the lower one at the speed $v=2.8$ mm/s. This speed corresponds to a Capillary number Ca$=\eta_0 v/\gamma$ smaller than 5$\times$10$^{-2}$, and, therefore, dynamical effects of the stretching process were expected to be small \citep{MP20}. To reduce the current crossing the liquid bridge (and the associated Joule effect), a small electric potential was set at the beginning of the experiment. When the electric current measured by the power supply fell below a threshold, the applied voltage was instantaneously increased up to its prescribed value. The images acquired in the experiments were processed with a sub-pixel resolution technique \citep{VMF11} to precisely determine the free surface position. We calculated the liquid bridge minimum diameter $d_{\textin{min}}$ from the free surface contour detected in the images. A quasi-cylindrical filament formed between the upper and parent drops during most of the liquid bridge breakup. The filament length $\ell_f$ was calculated as the distance between the two liquid bridge sections whose diameters were twice the minimum diameter.

% Possible errors
Water evaporation, enhanced by the heating caused by the Joule effect, can significantly change the polymer concentration in the course of the experiment. To rule out this possibility, we measured the liquid bridge volume during the experiment and verified that it did not change significantly. We also examined the images to discard the existence of visible bubbles. We verified that the capacitance of the electrodes had negligible effects by checking that the RC time constant was much smaller than any characteristic time of the problem. To ensure that the power supply, the camera and the picoammeter were correctly synchronized, we conducted an experiment with a liquid bridge of water and verified that the instant at which the current intensity vanished coincided with the free surface pinching.

\subsection{Tested liquids}

% Liquids
The fluids used in the experiments were polymeric solutions in deionized water (DIW) and a glycerin-water (G/W) mixture 50/50\% (w/w). In the two cases, the polymer was polyethylene oxide $2\times 10^6$ g/mol in molecular weight (PEO2M) ({\sc Sigma Aldrich}) at a concentration of 1\% (w). We will refer to these two solutions as DIW-PEO2M and G/W-PEO2M. Stock solutions were prepared by dissolving the polymers in the solvent with a magnetic stirrer at low angular speeds to minimize mechanical degradation of the long polymer chains. Additionally, and to avoid any other source of degradation, all the solutions were kept in a refrigerator prior to their use and covered with aluminium foil, which has reflective properties and acts as a barrier to light, preventing UV degradation of the polymeric molecules and microorganism growth.

% Properties
The density $\rho$ of the tested fluids was measured with a pycnometer of $5\pm 0.03$ ml and a precision balance. The solvent viscosity value $\eta^{(s)}$ was taken from Ref.\ \citep{GPA63}. The surface tension $\gamma$ was measured with the Theoretical Interface Fitting Analysis (TIFA) method \citep{CBMN04}. The extensional relaxation time $\lambda_e$ was measured with a CaBER rheometer \citep{MT00}. The specific heat capacity $c$ was taken from the literature \citep{GPA63}. Finally, the relative electrical permittivity $\beta$ was supposed to be that of the solvent, which was taken from Ref.\ \citep{GPA63}. Table \ref{t1} shows the properties of the tested liquids.

\begin{table}
\begin{tabular}{|c|c|c|c|c|c|c|c|}
\hline
&$\rho$ (kg/m$^3$)&$\eta^{(s)}$ (mPa$\cdot$s)&$\eta_0$ (Pa$\cdot$s)&$\gamma$ (mN/m)&$\lambda_e$ (ms)&$c$ (J/kg K)&$\beta$\\
\hline
DIW-PEO2M&$992.6\pm 0.1$&1.0&$0.088\pm 0.001$&$61\pm 1$&$13.0\pm 0.8$&$4179$&80\\
\hline
G/W-PEO2M&$1123.0\pm 0.1$&6.0&$0.931\pm 0.005$&$55\pm 1$&$11.3\pm 0.8$&3347&69\\
\hline
\end{tabular}
\caption{Properties of the tested liquids at 20 $^{\circ}$C.}
\label{t1}
\end{table}

\begin{figure}
\begin{center}
\includegraphics[width=0.35\linewidth]{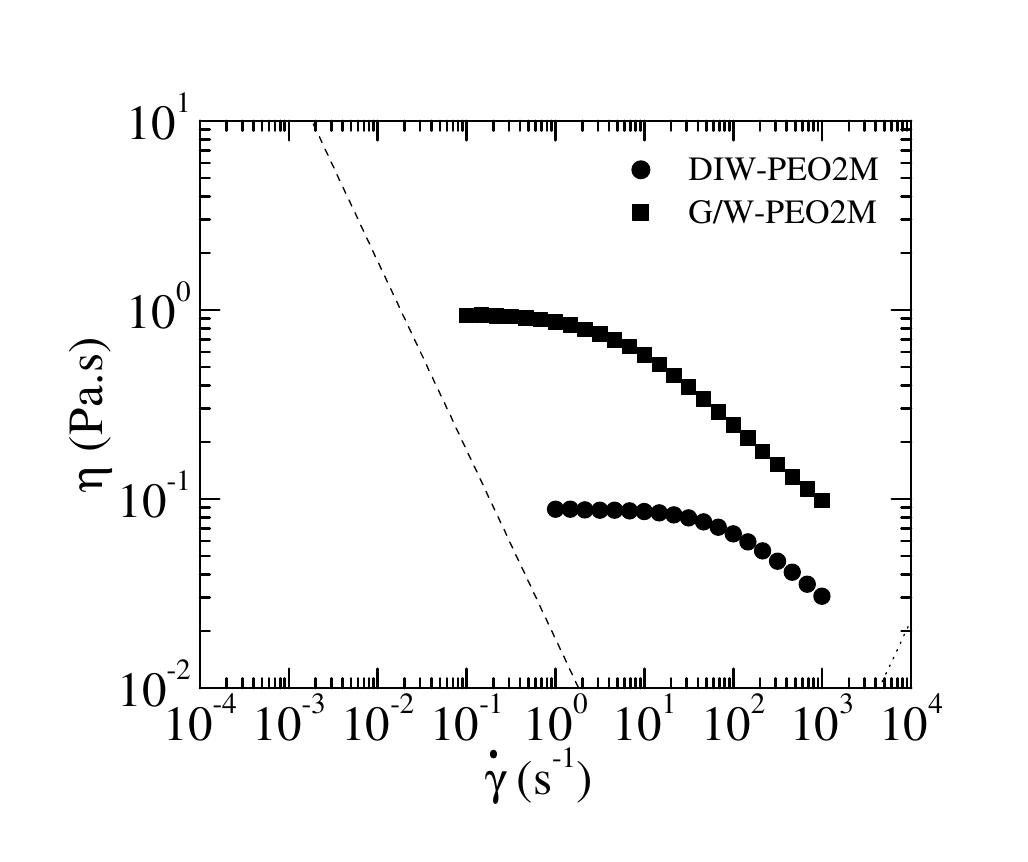}
\end{center}
\caption{Dependence of the solution shear viscosity $\eta$ upon the shear rate $\dot{\gamma}$ at 22 $^{\circ}$C. The dashed and dotted lines correspond to the low-shear-rate and elastic instability limits, respectively.}
\label{rheology}
\end{figure}

% Shear rheology
A stress controlled rotational rheometer (Anton Paar MCR301) was used to obtain the steady shear viscosity $\eta$ as a function of the shear rate $\dot{\gamma}$ of the fluid samples (Fig.\ \ref{rheology}). We used a cone-plate geometry of $R_c=37.5$ mm in radius with a cone angle of 1$^{\circ}$. The temperature within the fluid volume was set at 22 $^{\circ}$C and controlled by a Peltier element. Steady-state viscosity curves were obtained from 0.1 to 1000 s$^{-1}$. At least three independent measurements were performed to ensure the reproducibility of the results. The range of shear rate providing reliable data was set for each sample between the limit of the rheometer sensitivity (low-shear rate limit) and the onset of elastic instabilities (high-shear rate limit). The line in Fig.\ \ref{rheology} corresponding to the rheometer sensitivity limit was calculated using the expression $\eta_{\textin{min}}=3M_{0}/(2\pi R_c^{3}\dot\gamma)$, where $M_{0}$ is 20 times the torque resolution of the equipment ($10^{-7}$  N$\cdot$m) \citep{EJC15}. Elastic instabilities are assumed to become noticeable at the critical value Wi$_{\text{crit}}=100$ of the Weissenberg number Wi$=\lambda_e\dot{\gamma}$ \citep{EJC15}. The line in Fig.\ \ref{rheology} corresponding to the elastic instabilities was calculated using that criterion. As can be observed, the viscoelastic solutions exhibit considerable shear thinning. The zero shear viscosity $\eta_0$ is taken as the value for the smallest shear rate.

\begin{figure}
\begin{center}
\includegraphics[width=0.35\linewidth]{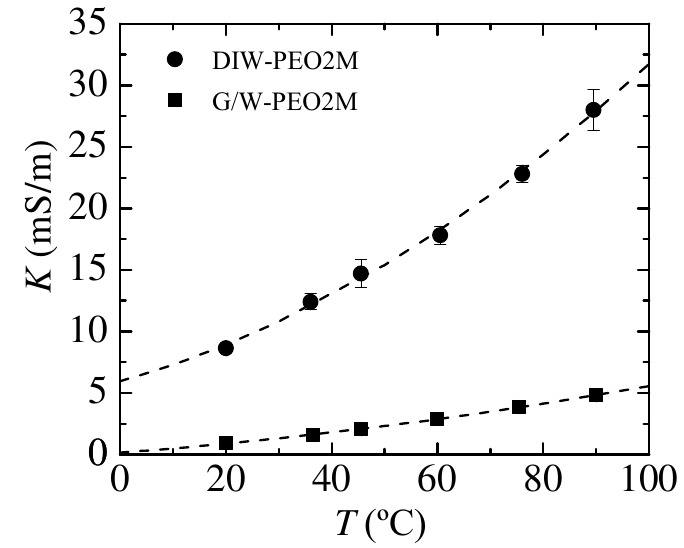}
\end{center}
\caption{Electrical conductivity as a function of the liquid temperature. The dashed lines are the second-degree polynomial fits to the experimental data.}
\label{cond}
\end{figure}

% Electrical conductivity
The dependency of the electrical conductivity upon the liquid temperature, $K(T)$, was measured with the following procedure. A cylindrical borosilicate capillary was submerged in a silicone oil bath kept at a fixed temperature controlled with a thermocouple. The capillary was filled with the tested liquid. We waited until the thermal equilibrium between that liquid and the surrounded bath was established. Then, a voltage difference was applied between the ends of the capillary, and the resulting electric current was measured. This measurement was conducted for several voltages in the range 5-30 V. The electrical conductivity was determined from the slope of the linear relationship between the applied voltage and measured electric current. Figure \ref{cond} shows the conductivity as a function of the temperature for the working liquids. Hereafter, $K(T)$ is referred to as the hydrostatic electrical conductivity.

\subsection{Measurement of the filament conductivity and extensional relaxation time}

% Charge conservation and cylindrical filament
A quasi-cylindrical liquid filament connects the upper and lower parent drops in most part of the liquid bridge breakup. The voltage drop across the parent drops is negligible compared to that taking place in the filament. Charge conservation as applied to the circuit sketched in Fig.\ \ref{sketch} leads to the following equations:
\begin{equation}
\label{e00}
I=\frac{V_f}{\tilde{R}_f}=\frac{V_o-I \tilde{R}_r}{\tilde{R}_f},
\end{equation}
where $I$ is the current intensity crossing the circuit, $V_o$ and $V_f$ are the applied electric potential and the voltage drop in the filament, respectively, while $\tilde{R}_r$ and $\tilde{R}_f$ are the electrical resistance of the resistor and filament, respectively. The filament diameter is practically the same as the minimum diameter $d_{\textin{min}}$ of the liquid bridge. Therefore, $\tilde{R}_f=4\hat{{\cal R}}_f\ell_f/(\pi d_{\textin{min}}^2)$, where $\hat{{\cal R}}_f$ is the filament resistivity (the inverse of the conductivity). Then, the resistivity $\hat{{\cal R}}_f$ is given by the expression
\begin{equation}
\label{e0}
\hat{{\cal R}}_f=\frac{V_o-I \tilde{R}_r}{I}\, \frac{\pi d_{\textin{min}}^2}{4\ell_f}.
\end{equation}
We will compare the values of the filament conductivity $K_f=\hat{{\cal R}}_f^{-1}$ obtained in the experiments with those of the conductivity $K(T)$ measured in hydrostatics with the standard procedure described above (Fig.\ \ref{cond}).

% Elasto-capillary regime
In the elasto-capillary regime, the minimum diameter $d_{\textin{min}}$ of the liquid filament decreases according to the exponential law
\begin{equation}
\label{dd}
d_{\textin{min}}=d_{\textin{min}0}\exp\left[-\frac{(t-t_0)}{3\lambda_e}\right],
\end{equation}
where $d_{\textin{min}0}$ is the minimum diameter at a reference instance $t=t_0$. To calculate $\lambda_e$, we fitted (\ref{dd}) to the experimental values of $d_{\textin{min}}(t)$ over a time interval of about $3\lambda_e$ at the beginning of the exponential decay.

% Joule effect
The filament temperature $T_f(t)$ increases over time due to Joule effect. If we neglect all kind of heat loss during the elasto-capillary regime, the filament temperature can be approximately calculated from the expression
\begin{equation}
\label{ttj}
T_f(t)\simeq T_{f0}+\int_{t_0}^t \frac{4I\, V_f}{\rho \pi d_{\textin{min}}^2\ell_{f} c}\, dt,
\end{equation}
where $T_{f0}$ is the temperature at the initial instant $t=t_0$ of the elasto-capillary regime, and specific heat capacity $c$ is assumed to be constant.

\section{Results}
\label{sec4}

\subsection{Numerical results}

% Introduction
The numerical and experimental problems considered in this work do not correspond to the same configuration. In the numerical analysis, the flow is supposed to be isotherm, and, therefore, the liquid physical properties are assumed to take constant values. On the contrary, the heating caused by the Joule effect in the experiments can lead to significant variations of the viscosity, conductivity, and surface tension, which entails the appearence of effects such as Marangoni convection. The maximum voltage leading to a stable solution of our model was 500 V, which is significantly different from the voltages of electrospinning considered in our experiments. This stems from the fact that we had to simplify the actual electrical boundary conditions in the simulation by considering two parallel large electrodes, as usually done in this problem. There are other less significant differences as well. In the theoretical model, surface conduction is produced only by the excess of charge, and the ion mobility is taken as the average value between the mobilities of H$^+$ and OH$^-$ in water. In the experiments, impurities coming from the atmosphere can be adsorbed and ionized at the free surface, and the ion mobility may significantly differ from that considered in the simulation. Finally, while the supporting disks remain static in the mathematical model, the upper disk moved up during the whole process. Despite all these approximations, the numerical solution of the leaky-dielectric FENE-P model helps us to interpret the experimental results. Specifically, we will show that (i) the electrical conduction measured in the experiments is due to the bulk conductivity exclusively (surface conduction is negligibe), and (ii) Maxwell stresses do not significantly alter the elasto-capillary balance of stresses over the filament thinning. 

% Parameters
In this section, we describe the numerical results calculated for $\hat{{\cal V}}=1$, $\Lambda=1.11$, $B=0.8$, $\lambda_s=0.939$, Oh$_0=2.65$, Oh$_s=0.0171$, $L^2=50104$ \citep{CPKOMSVM06}, $\alpha=1.3\times 10^6$, $\lambda=\pm 3.21$ (the sign $+/-$ applies to positive/negative charges), $\beta=69$, $R_o=5$, and $\chi=0.02$. This set of parameters corresponds to G/W-PEO2M and $V_o=500$ V.

\begin{figure}
\begin{center}
\includegraphics[width=0.35\linewidth]{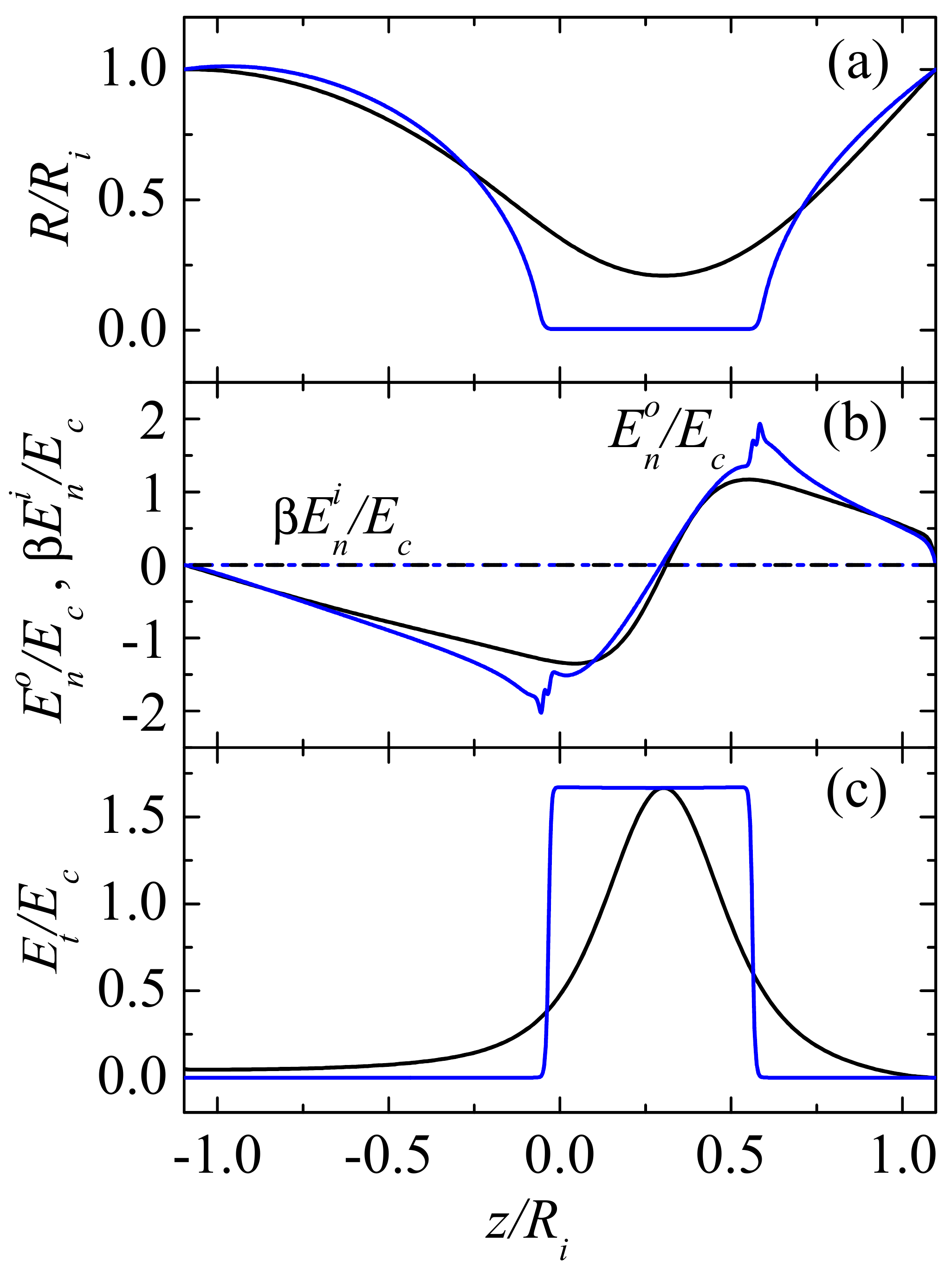}
\end{center}
\caption{Free surface contour $R$ (a), inner $E_n^i$ and outer $E_n^o$ normal components of the electric field (b), and tangential component $E_t$ of the electric field (c). The black and blue lines correspond to the instants $t/t_c=46.83$ and $t/t_c=77.38$, respectively.}
\label{er}
\end{figure}

% The characteristic times 
The dynamical process undergone by the liquid bridge consists of the inertio-capillary and elasto-capillary stages. Due to the moderately large liquid viscosity, the characteristic time of the inertio-capillary phase is larger than the capillary time $t_c$. In our simulation, the dimensionless conductivity $\alpha=\beta t_c/t_e$ takes a very large value, and, therefore, the electric relaxation time $t_e=\beta\varepsilon_0/K$ is much smaller than $t_c$. This means that the surface charge density almost relaxes to the electrostatic distribution at any time. As a result, the inner normal electric field takes very small values over the entire process. Figure \ref{er} shows the electric field over the free surface right after the voltage drop is applied ($t/t_c=46.83$) and once the elasto-capillary regime has been established ($t/t_c=77.38$). 

\begin{figure}
\begin{center}
\includegraphics[width=0.45\linewidth]{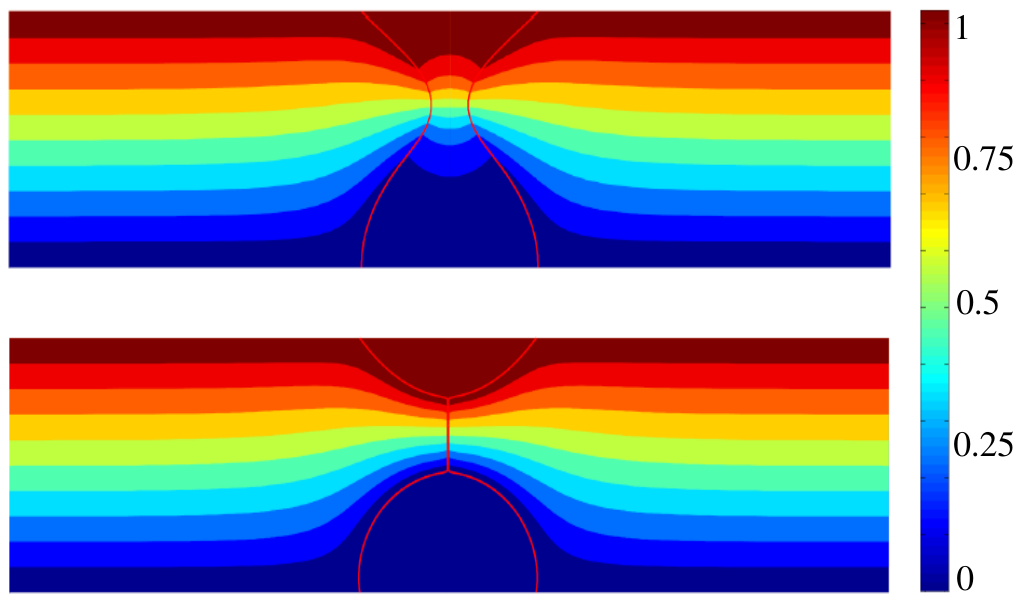}
\end{center}
\caption{Isolines of electric potential at the instants $t/t_c=46.83$ (upper image) and $t/t_c=77.38$ (lower image). The color scale indicates the values of the dimensionless electric potential $\phi ^{i,o}/V_0$.}
\label{potential}
\end{figure}

% Electric fields
In the elasto-capillary regime, a quasi-cylindrical filament is formed. Due to the large difference between the cross-sectional areas of the filament and the rest of the liquid bridge, the drop of voltage takes place almost entirely in the filament, while the parent drops remain at a constant potential. The first-order approximation of the electric field in the cylindrical filament is $E_z^i\simeq V_0/\ell_f$ and $E_r^i\simeq 0$. The outer radial electric field vanishes at the center of the filament on account of symmetry. In a significant part of the filament, $E_r^o$ exhibits a linear dependence with respect to the distance from the center, with a slope fixed by the instantaneous shape of the entire liquid bridge (Fig.\ \ref{er}). The electric field on the free surface substantially differs from that observed in cylindrical liquid bridges. In such configurations, the electric field is perfectly aligned with the free surface, and, therefore, both the normal electric field and the surface charge density vanish \citep{BS02}. In our simulation, the parent drops formed during the liquid bridge breakup considerably alter the axial electric field imposed by the electrodes (Fig.\ \ref{potential}), which makes the outer normal electric field be commensurate with the tangential one (Fig.\ \ref{er}). Thus, one can say that the surface charge density $\sigma\simeq \varepsilon_o E_r^o$ in the filament is somehow determined by what occurs outside that region. This resembles what happens to the tensile force, which is essentially built up in the corners of the filament.

\begin{figure}
\begin{center}
\includegraphics[width=0.69\linewidth]{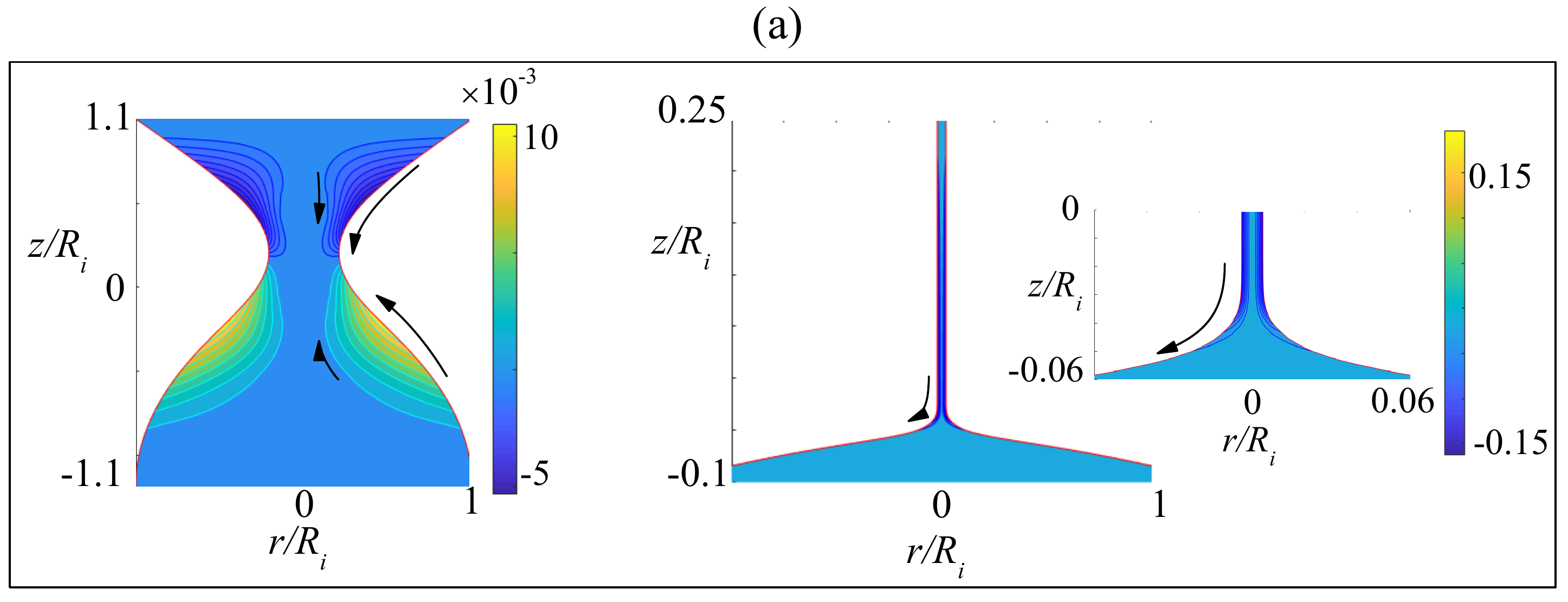}\includegraphics[width=0.31\linewidth]{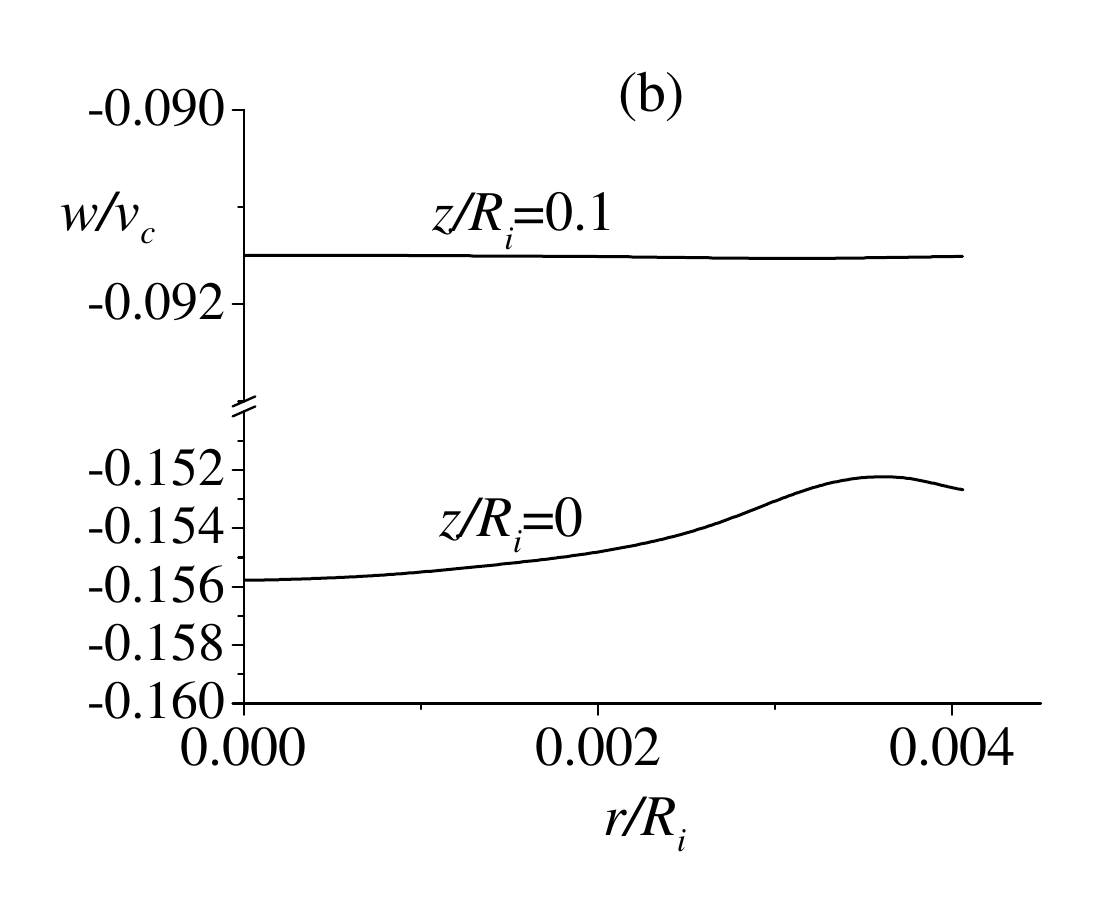}
\end{center}
\caption{(a) Streamlines at the instants $t/t_c=46.83$ (left) and $t/t_c=77.38$ (right). The color scales indicate the value of the stream function divided by $v_c^2 R_i$. The arrows indicate the flow direction. The right-hand graph shows a zoomed in view of the filament end. (b) Axial velocity profile $w(r)$ at the instant $t/t_c=77.38$.}
\label{streamlines}
\end{figure}

% The recirculation pattern
Figure \ref{streamlines} shows the streamlines right after the voltage is applied ($t/t_c=46.83$) and once the elasto-capillary regime has been established ($t/t_c=77.38$). The liquid flows towards the bridge neck during a very short time following the voltage switching. Then, it evacuates the central part of the liquid bridge to form the elasto-capillary filament. The charge accumulated in the interface during the elasto-capillary regime produces a shear electric stress $\sigma E_t$ which might feed recirculation cells in the liquid bridge, as occurs in electrospinning \citep{BHGM19}. Equation (\ref{int2}) allows us to estimate the characteristic velocity $v_{\textin{rc}}$ of these cells in terms of the inertio-capillary velocity $v_c=R_i/t_c$. Assuming that $v_{\textin{rc}}\sim w_r R_i$, the balance of shear stresses, $\sigma E_t\sim \mu w_r$, yields $v_{\textin{rc}}/v_c\sim (\sigma E_t/p_c)/\text{Oh}_0\sim 10^{-3}$, which explains the absence of recirculation cells in the simulation (Fig.\ \ref{streamlines}a). At $t/t_c=77.38$, the shear electric stress slightly bends the axial velocity profile next to the filament end ($z/R_i\simeq 0$) (Fig.\ \ref{streamlines}b), where the magnitude of $E_n^o$ reaches its maximum value (Fig.\ \ref{er}).

\begin{figure}
\begin{center}
\includegraphics[width=0.375\linewidth]{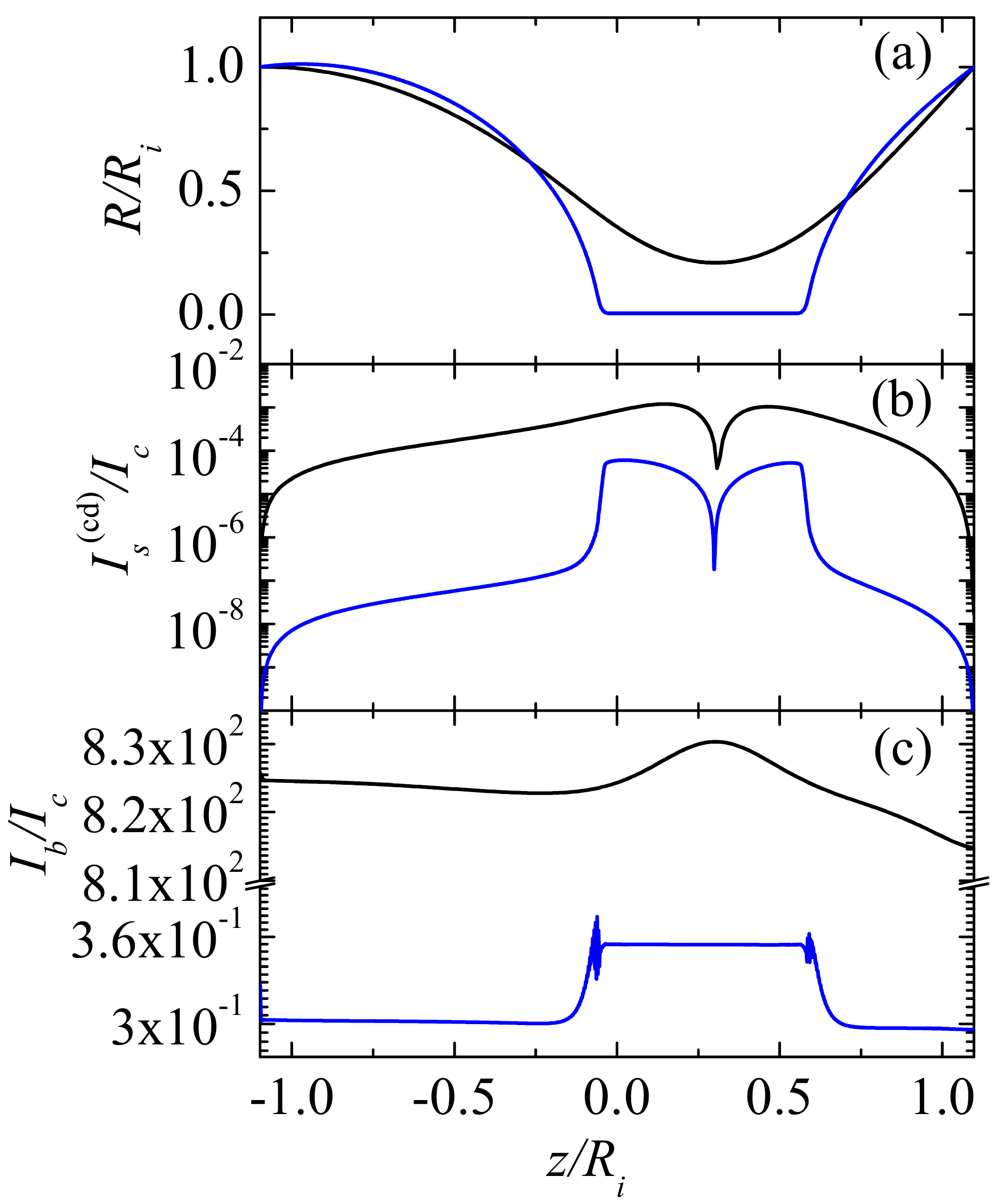}
\end{center}
\caption{Free surface contour $R(z)$ (a), and magnitude of surface conduction, $I_s^{(\textin{cd})}(z)$ (b), and bulk Ohmic conduction, $I_b(z)$ (c). The black and blue lines correspond to the instants $t/t_c=46.83$ and $t/t_c=77.38$, respectively. $I_c=[\gamma^3 R_i/(\rho V_0^2)]^{1/2}$ is the characteristic electric current.}
\label{inti}
\end{figure}

% The intensities
In principle, the total electric current between the two electrodes is the sum of bulk Ohmic conduction $I_b$, the surface conduction $I_s^{(\textin{cd})}$ quantified by the surface conductivity $\sigma\lambda$, and the surface convection $I_s^{(\textin{cv})}$ associated with the surface velocity [see Eq.\ (\ref{current})]. Due to the small surface velocity, surface convection of charge is negligible at any instant. Figure \ref{inti} compares the two conduction mechanisms. For $t/t_c=46.83$, surface conduction is many orders of magnitude smaller than the bulk one. As the liquid bridge approaches its breakup, the surface-to-volume ratio increases and surface conduction becomes more relevant. However, surface conduction remains several orders of magnitude smaller than the bulk one even for $t/t_c = 77.38$. It must be noted that the filament diameter corresponding to that instant is around 19 $\mu$m, smaller than those analyzed in our experiments. Therefore, we can conclude that surface conduction does not play any significant role in our experiments. It is worth noting that the Debye’s length is typically on the nanometer scale, much smaller than the filament diameter, which ensures the validity of the leaky-dielectric model \citep{GLHRM18}.

\begin{figure}
\begin{center}
\includegraphics[width=0.335\linewidth]{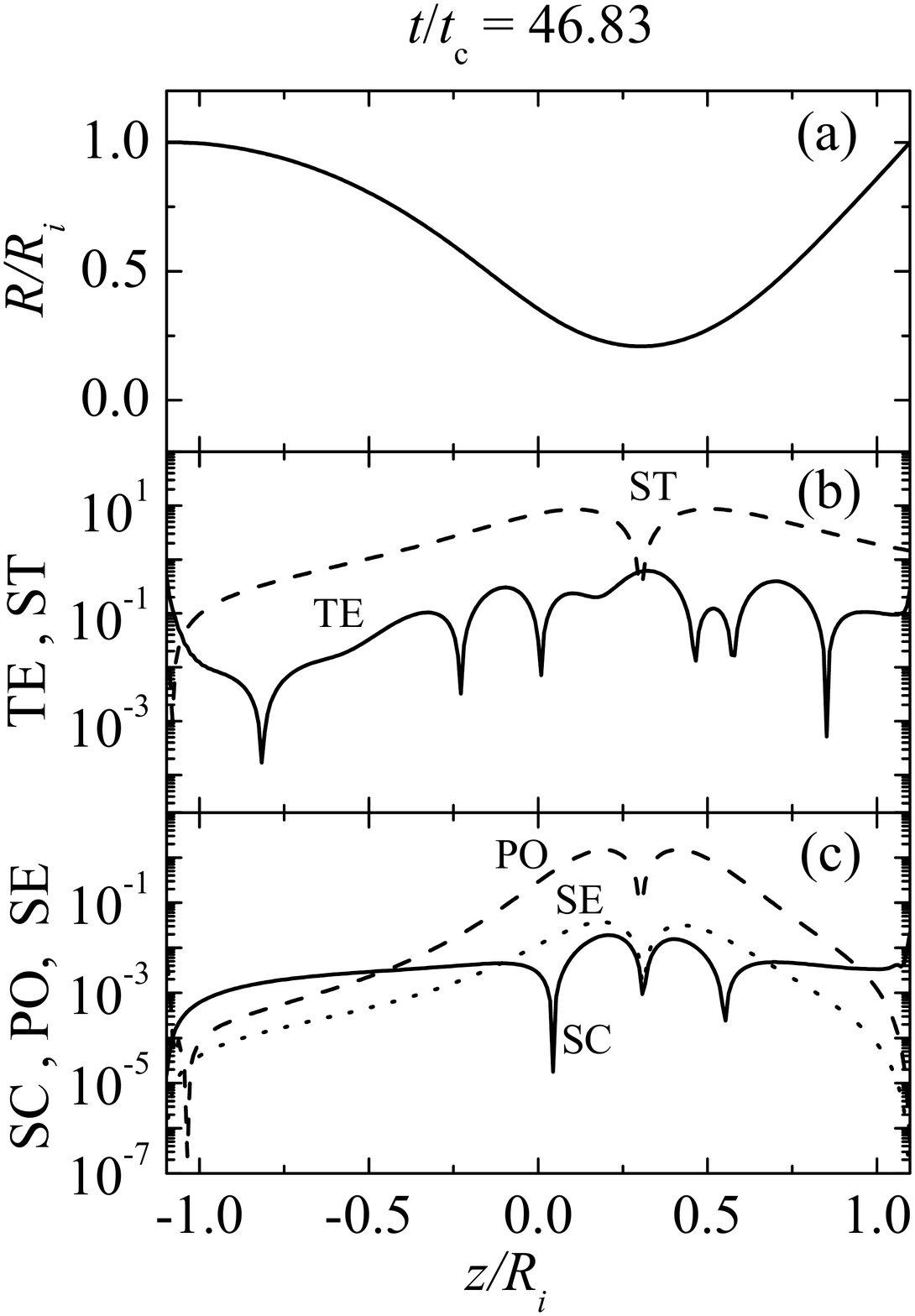}\hspace{0.5cm}\includegraphics[width=0.335\linewidth]{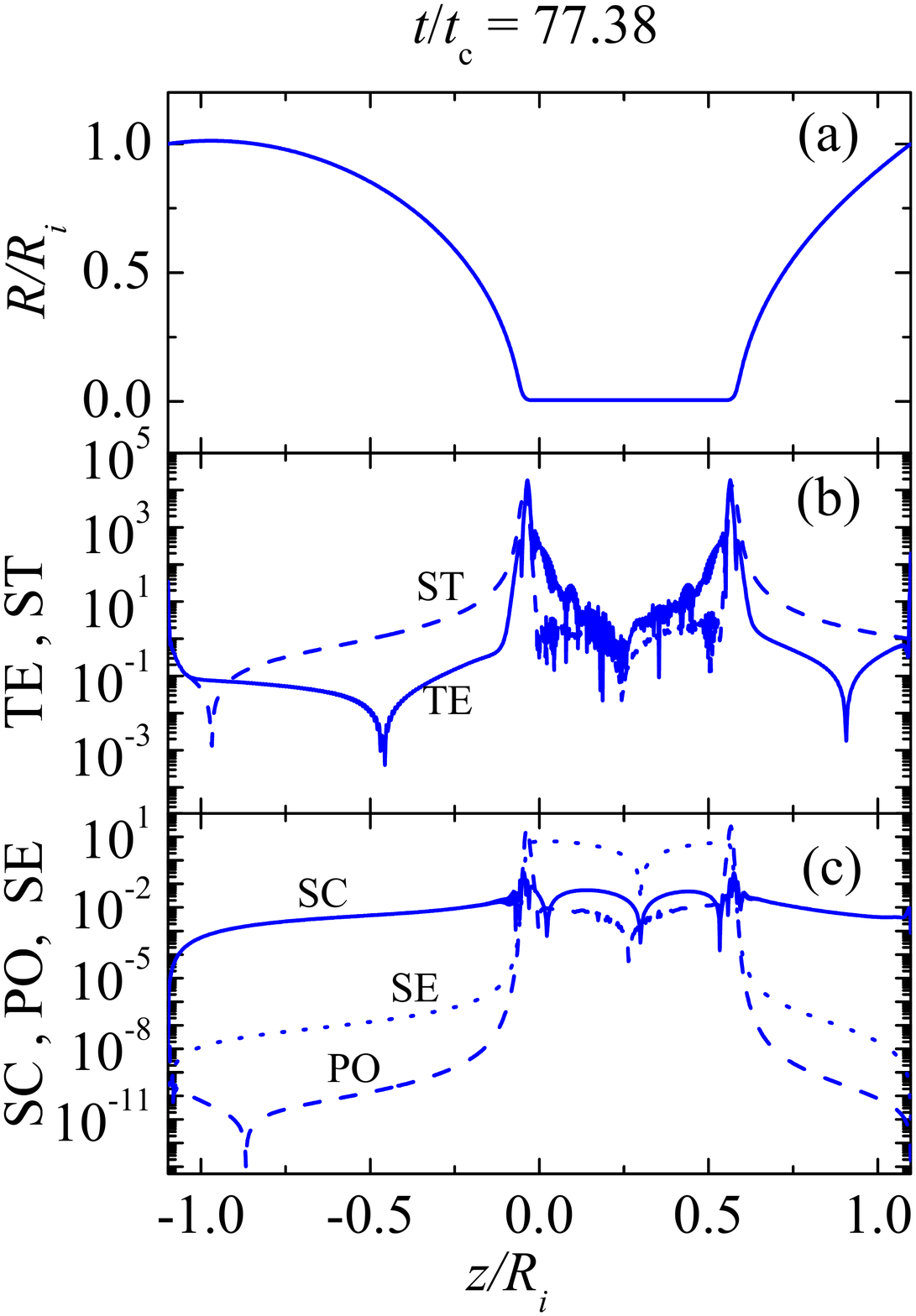}
\end{center}
\caption{Free surface contour $R(z)$ (a), and magnitude of the forces per unit volume exerted on a slice of the liquid bridge between $z$ and $z+dz$ (b,c). The labels in the panels (b) and (c) are those displayed in Eq.\ (\ref{mom}): tensile (TE) force, surface tension (ST) force, and electric force due surface charge (SC), polarization (PO) and shear electric (SE) stress. The forces have been made dimensionless with the characteristic for per unit volume $p_c/R_i$. The left-hand and right-hand graphs correspond to the instants $t/t_c=46.83$ and $t/t_c=77.38$, respectively. All the forces have been made dimensionless with $p_c/R_i$.}
\label{stresses}
\end{figure}

% Forces
Figure \ref{stresses} shows the magnitude of the forces per unit volume exerted on a slice of the liquid bridge between $z$ and $z+dz$. The labels are those displayed in Eq.\ (\ref{mom}). At $t/t_c=46.83$, the surface tension force (ST) dominates over both the viscoelastic (TE) and electrical ones. In this phase of the breakup, surface tension drives the motion and is essentially balanced by inertia. The polarization force (PO) is the most important one among the electrical forces due to the large value of the liquid permittivity. Surface tension pushes the liquid towards the parent drops emptying the filament which connects those drops. Elastic stresses grow as the filament thins, which gives rise to the elasto-capillary regime. At $t/t_c=77.38$, the surface tension is essentially balanced by the polymeric stress at the two ends of the cylindrical filament. The electrical forces are several orders of magnitude smaller than the viscoelastic and capillary ones in that region. The magnitude of the surface tension and viscoelastic forces sharply decreases in the cylindrical filament, where the surface charge and the intense axial electric field produce an intense shear electric force (SE). This force becomes comparable to and even greater than the surface tension and elasticity forces in the central part of the filament. This is in part because the area of the filament free surface (on which the electric shear stress is applied) becomes much larger than the cross-sectional area (on which the capillary and viecoelastic stress are applied) as the filament thins. It must be noted that this effect does not significantly affect the elasto-capillary balance (\ref{intmom}) in the filament, which is derived from the spatial integration from the upper/lower drop of the forces represented in Fig.\ \ref{stresses} \citep{EV08}. The major contribution to that integration comes from the elastic and capillary forces at the ends of the filament. It is worth mentioning that, under certain conditions, the electrical shear force enhances the asymmetric instability over the varicose mode in low-conductivity viscoelastic jets, which causes the bending motion in many experimental observations of electrospinning \citep{HSRB01a,XYQF17}. The polarization force is smaller than the shear electric force in the elasto-capillary regime. The electric pressure force (SC) is subdominant over the entire liquid bridge breakup. The (dimensionless) gravitational force equals -0.1, and is also subdominant except right in the middle of the filament. The fluctuations observed in the panels (b) and (c) of the right-hand graph can be attributed to the so-called high Weissenberg number instability \citep{FK04}, which produces numerical noise amplified by the axial derivative. 

% Electric Bond number
The electric Bond number defined in terms of the filament radius becomes of order unity at the end of the filament thinning, which might suggest that the electric pressure can affect the elasto-capillary balance established during the filament thinning. However, the magnitude of the normal electric field in the filament is not determined by the radial scale but by the filament length $\ell_f$ ($E_n^o\sim E_t\sim V_0/\ell_f$). Therefore, the electric Bond number measuring the relative importance of the electric normal stress to the capillary stress in the filament is $(\varepsilon_0 V_0^2/\ell_f^2)/(\gamma/d_{\textin{min}})=\chi (d_{\textin{min}}/\ell_f)^2$. This parameter takes values much smaller than unity because of the filament slenderness.

\begin{figure}
\begin{center}
\includegraphics[width=0.335\linewidth]{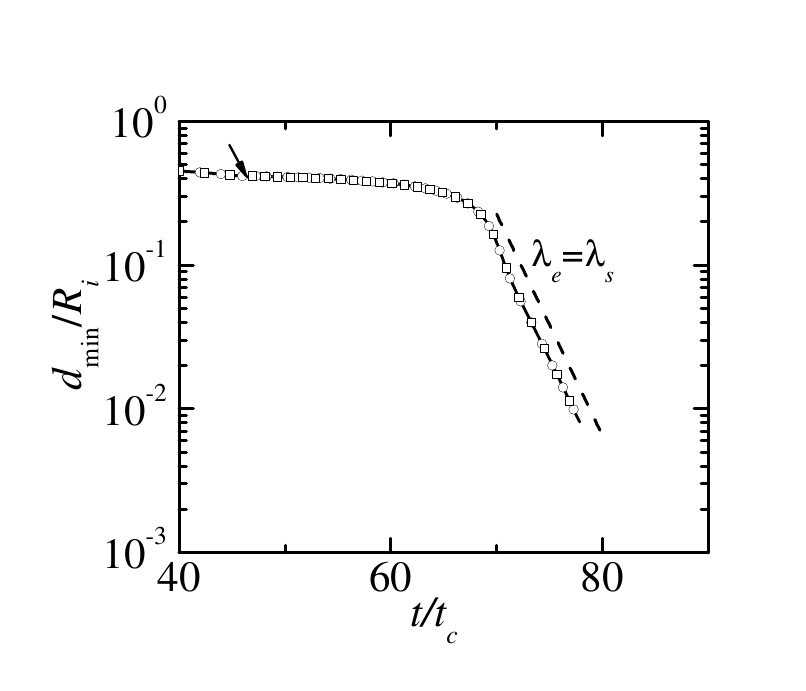}
\end{center}
\caption{Minimum diameter $d_{\textin{min}}$ as a function of time. The solid line and circles correspond to the solution to the FENE-P and Oldroyd-B models respectively. The dashed line corresponds to the exponential decay (\ref{dd}) with $\lambda_e=\lambda_s$. The arrow indicates the instant at which the voltage is applied. The squares corresponds to the solution to the FENE-P model for a non-electrified liquid bridge ($V_0=0$).}
\label{dimu}
\end{figure}

% Thinning
The results presented above indicate that Maxwell stresses do not alter the elasto-capillary balance during the exponential filament thinning, at least under conditions considered in our simulation. Therefore, the exponential relaxation time $\lambda_e$ measured from the time evolution of the minimum diameter $d_{\textin{min}}$ [Eq.\ (\ref{dd})] must coincide with the stress relaxation time $\lambda_s$ of the FENE-P model (\ref{h10}), as shown in Fig.\ \ref{dimu}. In other words, the possible effects of the electric field on $\lambda_e$ may be attributed to the influence of the electric field on the polymer stretching. The prefactor of the exponential thinning law in Fig.\ \ref{dimu} is much larger than the (dimensionless) elasto-capillary length $2(E_c/2)^{1/3}=2.28$, where $E_c=(\eta_0-\eta^{(s)}) R_i/(\lambda_s\gamma)$ is the elasto-capillary number \citep{CEFLM06,EHS20}. This is because the Deborah number $\lambda_s/t_c=0.939$ is much smaller than the time $t/t_c\sim 65$  needed to form the primary filament, and, therefore, the polymer relaxes during the formation of the filament.

% Oldroyd-B model and voltage
Figure \ref{dimu} also shows $d_{\textin{min}}(t)$ calculated with the Oldroyd-B \citep{O50,J09,TLEAD18} model. The results are essentially the same for two reasons: (i) $L^2$ takes a very large value for PEO2M, and, therefore, the finite extensibility effects contemplated in the FENE-P model do not come up over the time interval analyzed in the simulation; and (ii) shear thinning does not play any significant role in the extensional shear-free flow arising in the filament during the liquid bridge breakup. In fact, shear thinning is expected to affect the liquid dynamics in an extensional rheometer only next to the supporting disks, at begining of the liquid bridge stretching, and for sufficiently large stretching speeds. Finally, Fig.\ \ref{dimu} also shows $d_{\textin{min}}(t)$ calculated with the FENE-P model for a non-electrified liquid bridge ($V_0=0$). The electric stresses do not affect the evolution of the filament diameter.

\subsection{Experimental results}

\begin{figure}
\begin{center} 
\includegraphics[width=0.7\linewidth]{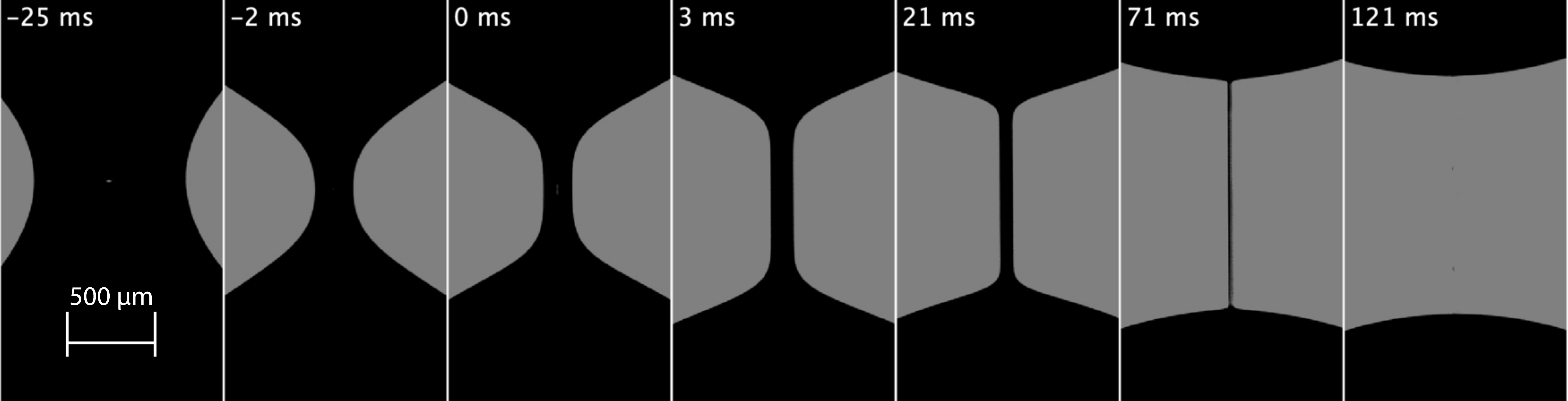}
\end{center}
\caption{Sequence of images of an experiment conducted with G/W-PEO2M, $V_o=1$ kV, and $\tilde{R}_r=1$ M$\Omega$. The origin of time is taken as the instant for which $d_{\textin{min}}=250$ $\mu$m.}
\label{images}
\end{figure}

\begin{figure}
\begin{center}
\includegraphics[width=0.35\linewidth]{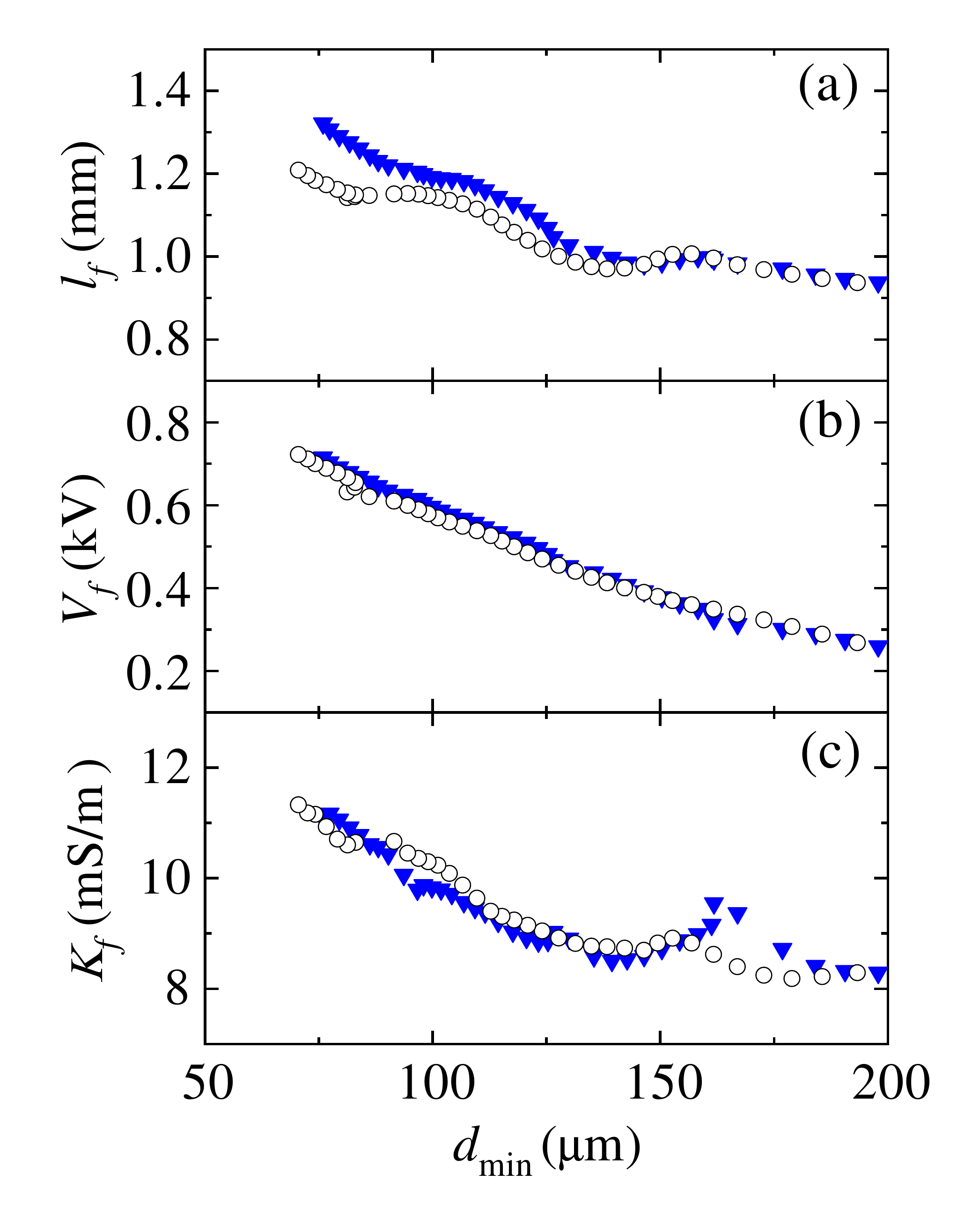}
\end{center}
\caption{Filament length $\ell_f$ (a), voltage $V_f$ (b), and filament conductivity $K_f$ (c) as a function of the minimum diameter $d_{\textin{min}}$ for DIW-PEO2M. The circles and triangles correspond to two experiments conducted for ($V_o=1$ kV, $\tilde{R}_r=1$ M$\Omega$).}
\label{ero}
\end{figure}

\begin{figure}
\begin{center}
\includegraphics[width=0.37\linewidth]{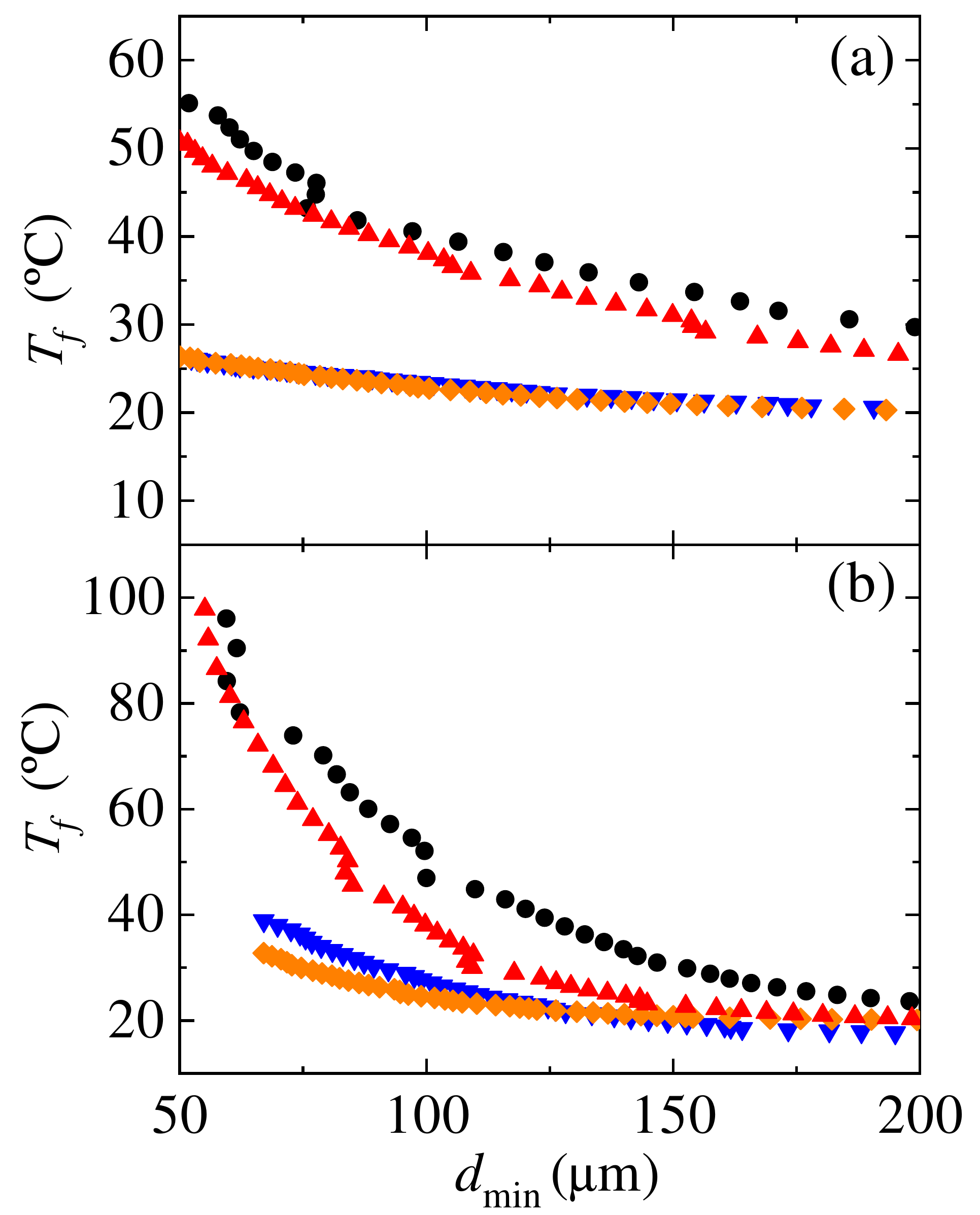}
\end{center}
\caption{Filament temperature $T_f$ as a function of the minimum diameter $d_{\textin{min}}$ for G/W-PEO2M (a) and DIW-PEO2M (b). The circles, up-triangles, down-triangles, and diamonds correspond to ($V_o=2$ kV, $\tilde{R}_f=1$ M$\Omega$), ($V_o=2$ kV, $\tilde{R}_r=2$ M$\Omega$), ($V_o=1$ kV, $\tilde{R}_r=1$ M$\Omega$), and ($V_o=1$ kV, $\tilde{R}_r=2$ M$\Omega$), respectively.}
\label{tt}
\end{figure}

\begin{figure}
\begin{center}
\includegraphics[width=0.37\linewidth]{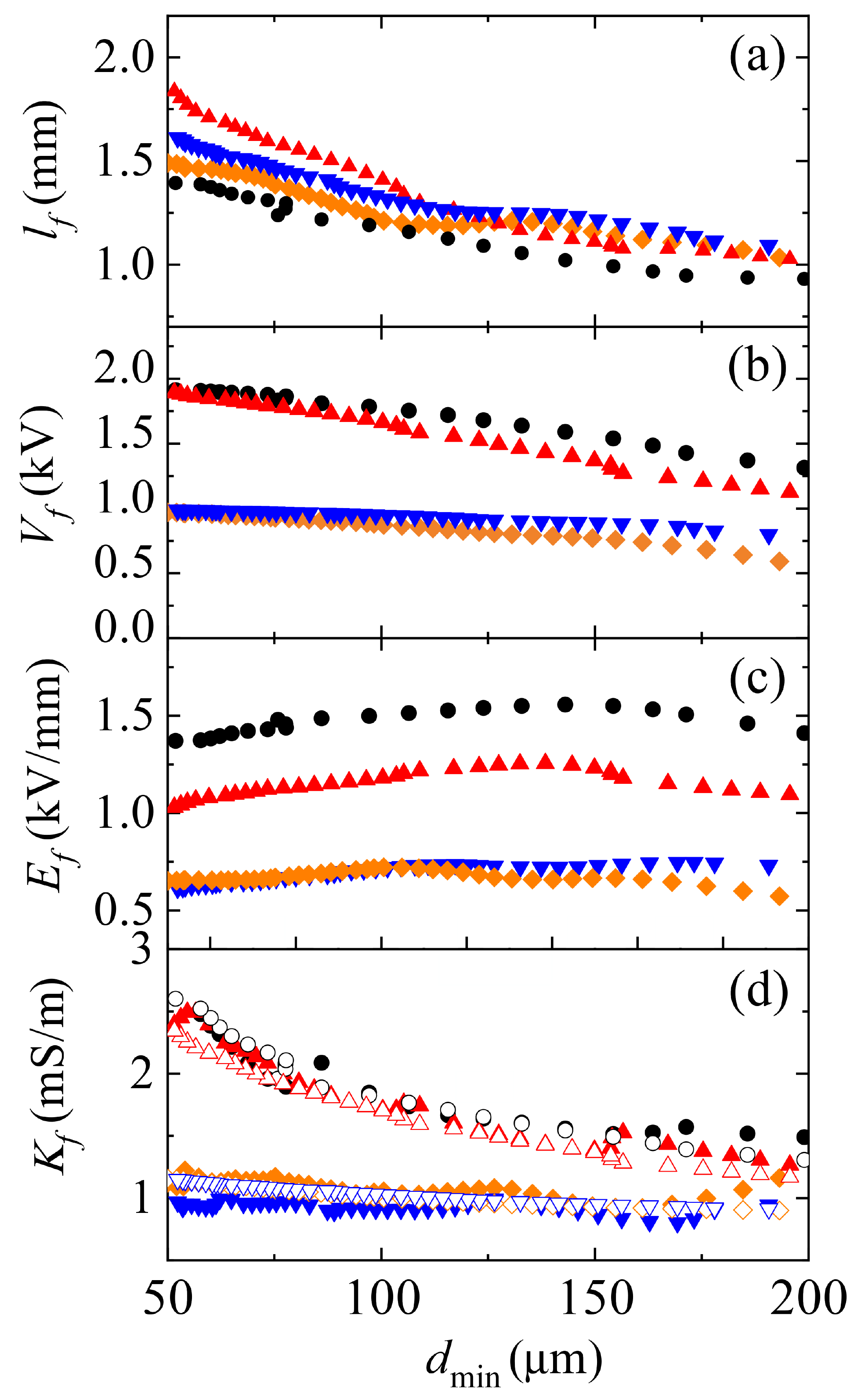}\includegraphics[width=0.37\linewidth]{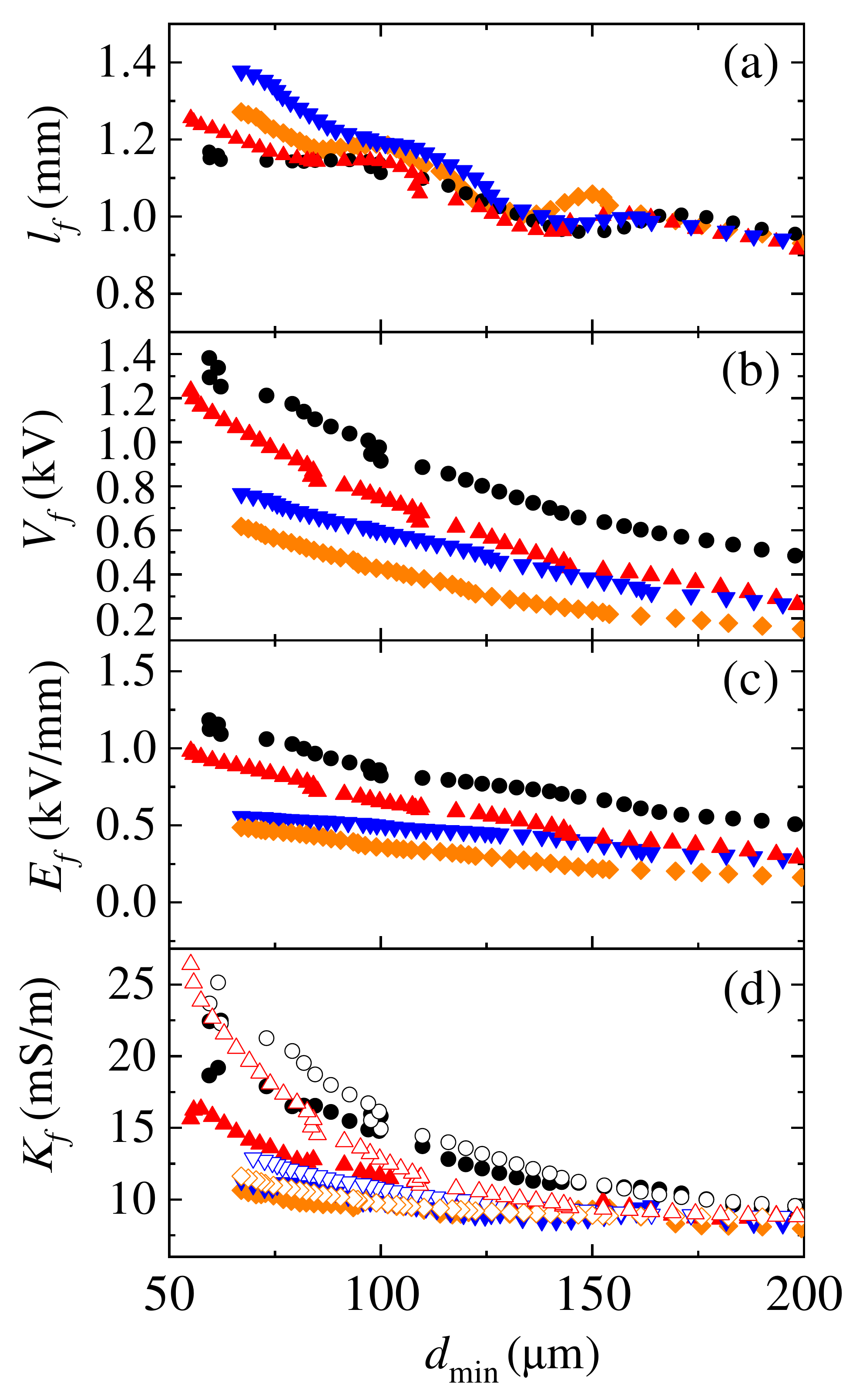}
\end{center}
\caption{Filament length $\ell_f$ (a), voltage drop $V_f$ (b), electric field $E_f$ (c), and conductivity $K_f$ (d) as a function of the minimum diameter $d_{\textin{min}}$ for G/W-PEO2M (left-hand panels) and DIW-PEO2M (right-hand panels). The circles, up-triangles, down-triangles, and diamonds correspond to ($V_o=2$ kV, $\tilde{R}_r=1$ M$\Omega$), ($V_o=2$ kV, $\tilde{R}_r=2$ M$\Omega$), ($V_o=1$ kV, $\tilde{R}_r=1$ M$\Omega$), and ($V_o=1$ kV, $\tilde{R}_r=2$ M$\Omega$), respectively. In the conductivity panel (d), the solid and open symbols correspond to the values measured in the course of the experiment and in hydrostatics, respectively.}
\label{DIW}
\end{figure}

The electrical conductivity of electrospun polymeric solutions is one of the key elements for the successful production of fibers in electrospinning and near-field electrospinning. One of the major motivations of this work is to determine the electrical conductivity of a thinning filament subject to voltage drops similar to those applied in electrospinning. To control the electric current in the course of the experiment, we varied not only the applied voltage $V_o$ but also the resistance $R_o$ of the resistor connected to the circuit (Fig.\ \ref{sketch}). We conducted experiments with G/W-PEO2M and DIW-PEO2M for $V_o=1$ and 2 kV, and for $R_o=1$ and 2 M$\Omega$. For the sake of illustration, Fig.\ \ref{images} shows a sequence of images acquired in one of our experiments. These images were analyzed and the electric current was measured to determine the electrical conductivity of the quasi-cylindrical filament formed between the two parent drops, as explained in Sec.\ \ref{sec3}. Filament diameters of the order of 10 $\mu$m and electric currents of the order 10-10$^2$ $\mu$A were measured at the end of the filament thinning. Despite the complexity of the problem, our experiments showed a high degree of reproducibility (Fig.\ \ref{ero}). The wavy shape of the curve $\ell_f(d_{\textin{min}})$ in Fig.\ \ref{ero} can be attributed to the criterion used to determine the filament length, which produces fluctuations of this quantity. These fluctuations do not affect our analysis. It is worth mentioning that the minimum diameter $d_{\textin{min}}$ decreases with time, and, therefore, the time evolution in Figs.\ \ref{ero}-\ref{DIW} must be read leftwards.

% Joule effect
As observed in Fig.\ \ref{images}, the thinning of the filament takes place over a time period of the order of 0.1 s. Despite the relatively large speed of this dynamical process, the Joule effect can produce a considerable increase in the filament temperature $T_f$. This occurs because the dissipated electrical energy is absorbed by the small volume of liquid trapped in the filament. It is worth mentioning that the residence time of a fluid particle in the heating region of electrospinning and near-field electrospinning may be much smaller than in the thinning filament of a liquid bridge. Therefore, liquid heating in electrospinning and near-field electrospinning may be considerably smaller than in our experiments.

Figure \ref{tt} shows the filament temperature calculated from Eq.\ (\ref{ttj}) for the four experiments considered in our analysis. As mentioned in Sec.\ \ref{sec3}, we did not consider any heat loss in our calculations, and, therefore, the filament temperature may be overestimated. In fact, we have verified that the evaporation of a small portion of the filament, not detectable in the experiment, would significantly decrease its temperature. Unfortunately, we cannot accurately calculate the evaporation rate because we did not control the ambient conditions. The filament temperature $T_{f0}$ at the initial instant was determined so that the hydrostatic value $K(T_{f0})$ (Fig.\ \ref{cond}) coincides with $K_f$ at that instant. 

% Results
Figure \ref{DIW} shows all the quantities measured in the experiments with both G/W-PEO2M and DIW-PEO2M. Both the filament length $\ell_f$ and the voltage drop $V_f$ across the filament slightly increases as the filament diameter decreases. As can be observed, the filament voltage $V_f$ increases as the applied voltage $V_o$ increases and/or the electrical resistance $R_o$ decreases. The axial electric field $E_f(t)=V_f(t)/\ell_f(t)$ in the filament remains practically constant for G/W-PEO2M, while it slightly increases during the thinning of the DIW-PEO2M filament. Overall, the filament conductivity increases as the diameter decreases. This effect becomes more noticeable as the voltage drop increases. The filament conductivity becomes up to three times its initial value for DIW-PEO2M and the largest voltage drop. 

% Comparison
Figure \ref{DIW} also shows the hydrostatic value $K(d_{\textin{min}})$ of the liquid conductivity obtained from the composition of the functions $K(T_f)$ and $T_f(d_{\textin{min}})$ represented in Figs.\ \ref{cond} and \ref{tt}, respectively. As can be observed, there is good agreement between the filament conductivity and the corresponding hydrostatic value for G/W-PEO2M. This suggests that the change of the filament microscopic structure due to the polymer stretching does not considerably affect the ion mobility in the stretching direction. On the contrary, the hydrostatic conductivity significantly exceeds the filament conductivity for DIW-PEO2M. However, this discrepancy may be attributed to the heat loss neglected in the calculation of the filament temperature, which leads to an overestimation of that temperature, and, therefore, of the hydrostatic conductivity. 

\begin{figure}
\begin{center}
\includegraphics[width=0.4\linewidth]{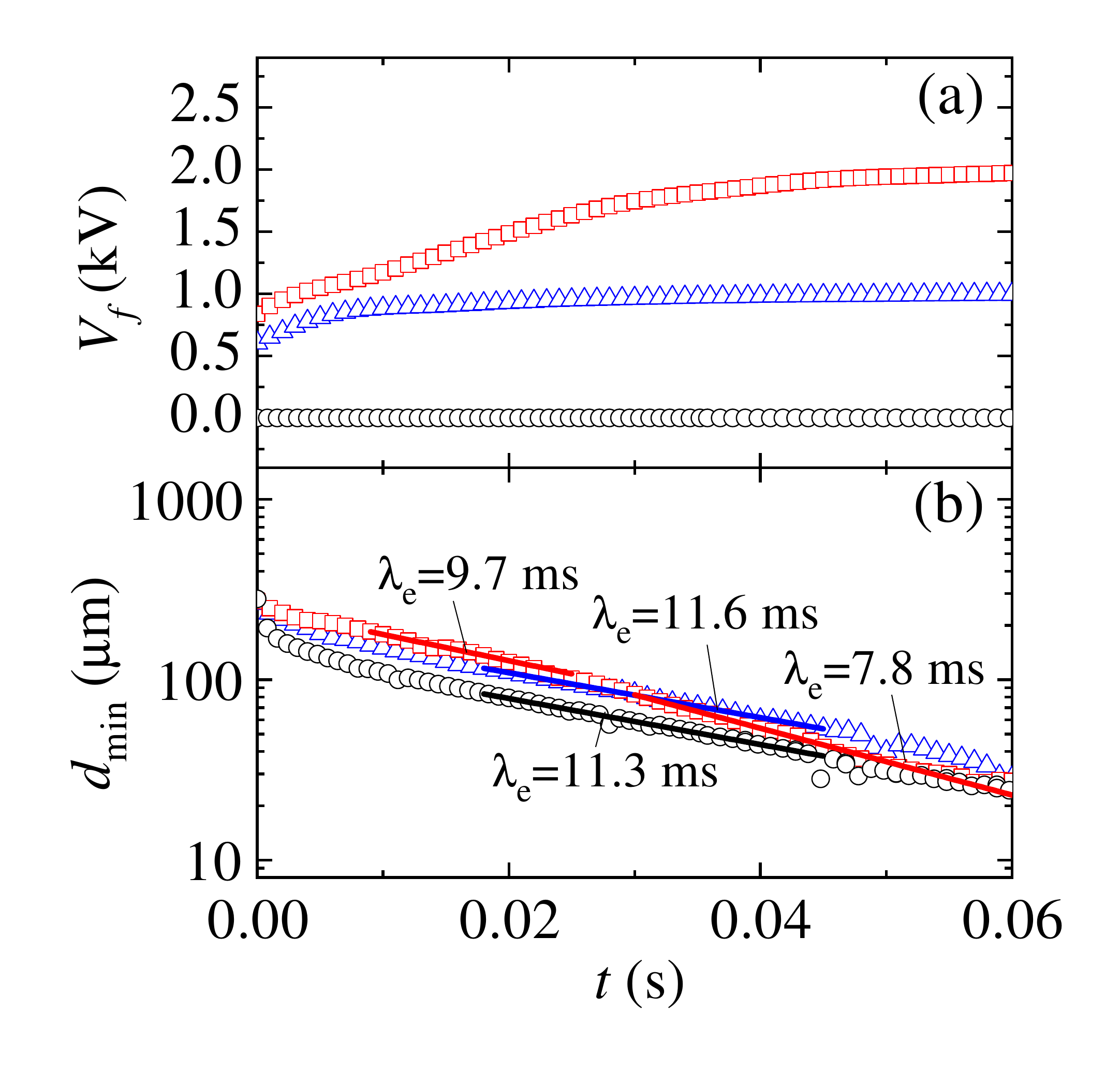}\includegraphics[width=0.39\linewidth]{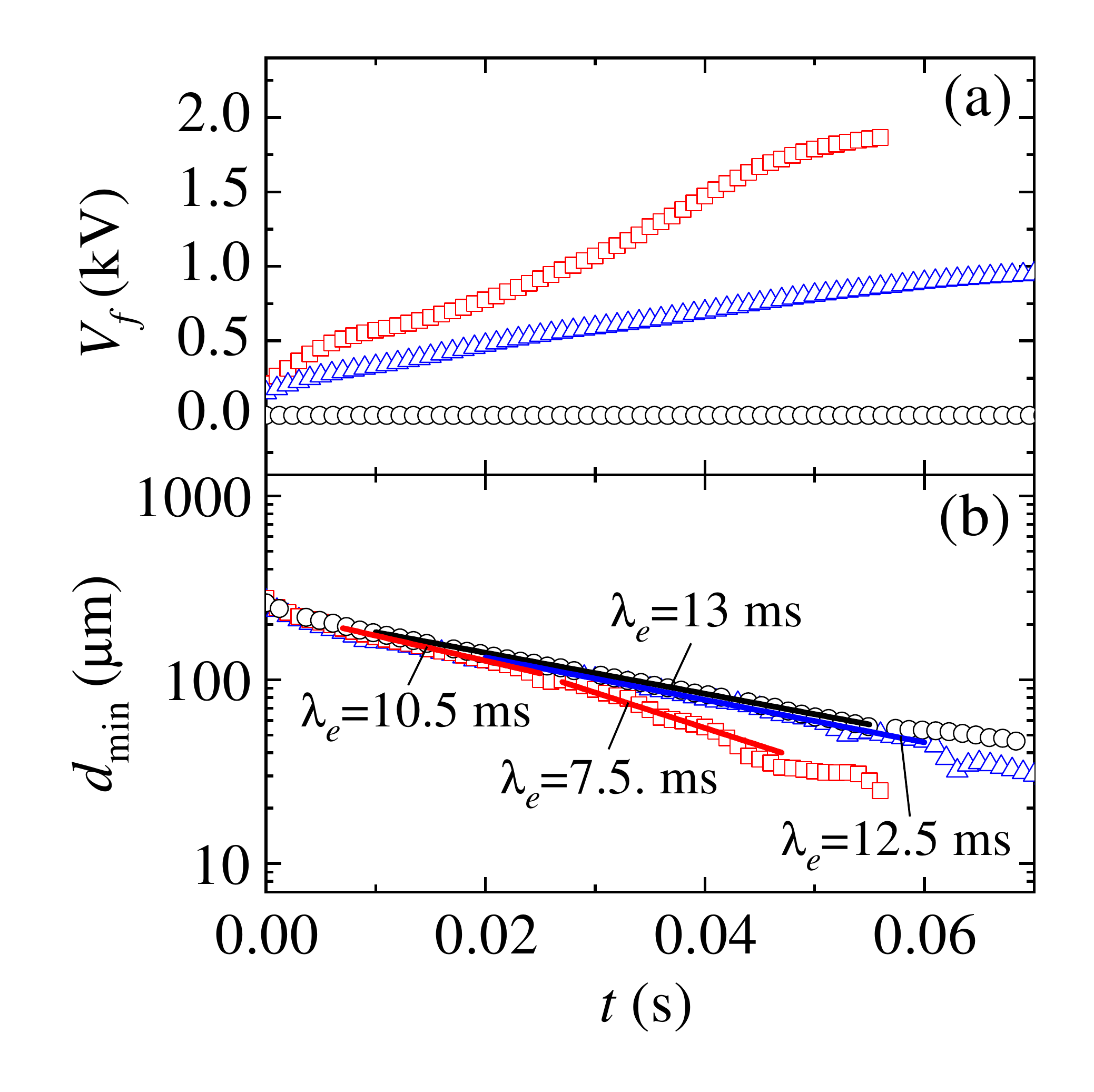}
\end{center}
\caption{Voltage $V_f$ (a) and minimum diameter $d_{\textin{min}}$ (b) as a function of time for G/W-PEO2M (left-hand panels) and DIW-PEO2M (right-hand panels). The circles, triangles and squares correspond to ($V_o=0$ kV, $\tilde{R}_r=1$ M$\Omega$), ($V_o=1$ kV, $\tilde{R}_r=1$ M$\Omega$), and ($V_o=2$ kV, $\tilde{R}_r=1$ M$\Omega$), respectively. The solid lines are the fit of (\ref{dd}) to the experimental data. The origin of time is taken as the instant for which $d_{\textin{min}}=250$ $\mu$m.}
\label{rt}
\end{figure}

% Extensional relaxation time
Figure \ref{rt} shows the temporal evolution of the filament minimum diameter $d_{\textin{min}}$ and the fits (\ref{dd}) to the experimental data. The relaxation times obtained in the first stage of the exponential thinning slightly depend on the applied voltage. However, for $V_o=2$ kV, this first stage yields a second phase where the exponential thinning is characterized by a smaller relaxation time. The difference between the two values of $\lambda_e$ is larger than the experimental uncertainty $\pm 0.8$ ms calculated as half of the maximum difference among the values obtained in 3 experimental realizations. As shown in the theoretical analysis, the extensional relaxation time is not significantly affected by the Maxwell stresses. Therefore, this decrease cannot be attributed to those stresses. A plausible cause could be the increase in the temperature during the filament thinning, which may alter the behavior of the polymer chains. This can be clearly observed in the case of DIW-PEO2M and ($V_{o}=2$ kV, $R_{r}=2$ M$\Omega$). In this case, the temperature sharply increases for  $d_{\textin{min}}\lesssim 100$ $\mu$m (Fig.\ \ref{tt}), which is the interval where the extensional relaxation time changes. Table \ref{ext} displays the values of $\lambda_e$ measured in all the experiments.

% Relaxation time

\begin{table}
\begin{tabular}{c|c|c|c|c}
\hline
Liquid&$V_o$ (kV)&$\tilde{R}_r$ (M$\Omega$)&\multicolumn{2}{c}{$\lambda_e$ (ms)}\\
\hline
G/W-PEO2M&0&1&\multicolumn{2}{c}{$11.3\pm 0.8$}\\
\hline
G/W-PEO2M&1&1&\multicolumn{2}{c}{$11.6\pm 0.8$}\\
\hline
G/W-PEO2M&2&1&$9.7\pm 0.8$&$7.8\pm 0.8$\\
\hline
DIW-PEO2M&0&1&\multicolumn{2}{c}{$13\pm 0.8$}\\
\hline
DIW-PEO2M&1&1&\multicolumn{2}{c}{$12.5\pm 0.8$}\\
\hline
DIW-PEO2M&2&1&$10.5\pm 0.8$&$7.5\pm 0.8$\\
\hline
\end{tabular}
\caption{Extensional relaxation time $\lambda_e$ measured in all the experiments.}
\label{ext}
\end{table}

\section{Conclusions}

% Results
We have studied both numerically and experimentally the breakup of a viscoelastic liquid bridge held by surface tension between two horizontal electrodes. The leaky-dielectric FENE-P model has been solved to examine the evolution of a liquid bridge under isothermal conditions, i.e. with constant physical properties (surface tension, viscosity, electrical conductivity, \ldots). The initial inertio-capillary stage of the breakup gives rise to the elasto-capillary regime, in which a thin liquid filament forms between the two parent drops. These drops are essentially at rest and at the same voltage as that of the electrode they are in contact with. Therefore, the voltage drops entirely in the liquid elasto-capillary filament, which produces an intense axial electric field in that region. The presence of the parent drops considerably alters the outer electric field. In fact, the outer electric field perpendicular to the interface is of the same order of magnitude as that of the axial electric field in the filament. On the contrary, the inner normal electric field practically vanishes at any time due to the fast transfer of electric charge from the bulk to the interface. The shear electric field caused by the accumulation of charges at the interface does not significantly affect the velocity field. The surface charge density is not large enough to contribute significantly to the total electric current over the filament thinning. One of the major conclusions of our analysis is that Maxwell stresses do not interfere in the measurement of the extensional relaxation time from the filament exponential thinning. In fact, those stresses play a negligible role from the beginning of the elasto-capillary thinning, and, therefore, one can safely identify the exponential relaxation time with the stress relaxation time in the FENE-P model.

We conducted experiments with polymer solutions and applied voltages similar to those commonly used in electrospinning and near-field electrospinning. The motivation was twofold: (i) to measure the electrical conductivity when the microscopic structure of the liquid is altered by the polymer stretching, and (ii) to measure the extensional relaxation time when the liquid is subject to a strong electric field. This information may be relevant to gain insight into the physical mechanisms governing electrospinning. The electrical conductivity of the thinning filament was compared with that measured in hydrostatic conditions for the same estimated temperature. Good agreement was found for G/W-PEO2M, which suggests that the stretching of the polymeric molecules does not significantly modify the ion mobility in the stretching direction. On the contrary, the conductivity of the DIW-PEO2M was considerably smaller than its hydrostatic counterpart. We speculate that the true filament temperature was smaller than that estimated without any heat loss, and, therefore, the hydrostatic conductivity was overestimated. We verified that the relaxation times obtained in the first stage of the exponential thinning hardly depend on the applied voltage. However, a small but measurable influence of the applied voltage was found in the last part of the time interval analyzed in the experiments.

% Concluding remarks

\vspace{1cm}
{\bf Acknowledgement.} This research has been supported by the Spanish Ministry of Economy, Industry and Competitiveness under Grant DPI2016-78887, and by Junta de Extremadura under Grant GR18175; and by Project PTDC/EME-APL/30765/2017 - POCI-01-0145-FEDER-030765 - funded by FEDER funds through COMPETE2020 - Programa Operacional Competitividade e Internacionalização (POCI) and with financial support of FCT/MCTES through national funds (PIDDAC).

% Bibliography
%\bibliography{central}\end{document}

%merlin.mbs apsrev4-1.bst 2010-07-25 4.21a (PWD, AO, DPC) hacked
%Control: key (0)
%Control: author (0) dotless jnrlst
%Control: editor formatted (1) identically to author
%Control: production of article title (0) allowed
%Control: page (1) range
%Control: year (0) verbatim
%Control: production of eprint (0) enabled
%

\end{document}